\documentclass[12pt,twoside,a4paper]{article}
\usepackage{amsmath,amssymb,latexsym,theorem,bbm,natbib,color}
\usepackage{multirow,graphicx}  
\usepackage{url}
\newfont{\msa}{msam10 scaled\magstep1}
\newfont{\ssmsa}{msam9}

\def\crps{\mathop{\hbox{\rm CRPS}}}
\def\logs{\mathop{\hbox{\rm LogS}}}
\def\md{\mathop{\hbox {\rm MD}}}

\numberwithin{equation}{section}

\title{Combining predictive distributions for statistical post-processing of ensemble forecasts}

\author{{\sc S\'andor Baran$^{1}$} and {\sc Sebastian Lerch$^{2,3}$} \vspace*{0.5cm}\\
         $^1$Faculty of Informatics, University of Debrecen\\
         Kassai \'ut 26, H-4028 Debrecen, Hungary \\
         $^2$ Heidelberg Institute for Theoretical Studies\\
         Schloss-Wolfsbrunnenweg 35, D-69118 Heidelberg, Germany \\
         $^3$ Institute for Stochastics, Karlsruhe Institute of Technology, \\
         Englerstra\ss{}e 2, D-76128 Karlsruhe, Germany}

\date{\today}

\begin{document}
\pagestyle{myheadings}

\maketitle

\begin{abstract}
Statistical post-processing techniques are now widely used to correct systematic biases and errors in calibration of ensemble forecasts obtained from multiple runs of numerical weather prediction models. A standard approach is the ensemble model output statistics (EMOS) method, a distributional regression approach where the forecast distribution is given by a single parametric law with parameters depending on the ensemble members. Choosing an appropriate parametric family for the weather variable of interest is a critical, however, often non-trivial task, and has been the focus of much recent research. In this article, we assess the merits of combining predictive distributions from multiple EMOS models based on different parametric families. In four case studies with wind speed and precipitation forecasts from two ensemble prediction systems, we study whether state of the art forecast combination methods are able to improve forecast skill.

\end{abstract}

\section{Introduction}
  \label{sec:sec1}
 
Nowadays, weather forecasts are typically based on the output of numerical weather prediction (NWP) models which describe the dynamical and physical behavior of the atmosphere through nonlinear partial differential equations. Single deterministic predictions produced by single runs of such models fail to account for uncertainties in the initial conditions and the numerical model. Therefore, NWP models are nowadays typically run several times with varying initial conditions and model physics, resulting in an ensemble of forecasts, see \citet{palmer2002} and \citet{gr05} for reviews. Examples of ensemble prediction systems (EPSs) are the 51-member European Centre for Medium-Range Weather Forecasts (ECMWF) ensemble \citep{ecmwfens}, the eight-member University of Washington Mesoscale ensemble \citep[UWME;][]{em05}, and the 11-member Aire Limit\'ee Adaptation dynamique D\'eveloppement International-Hungary Ensemble Prediction System \citep[ALADIN-HUNEPS;][]{hkkr} of the Hungarian Meteorological Service (HMS). The transition from single deterministic forecasts to ensemble predictions can be seen as an important step towards probabilistic forecasting, however, ensemble forecasts are often underdispersive and subject to systematic bias. They thus require some form of statistical post-processing.

Over the past decade, various statistical post-processing methods have been proposed in the meteorological literature. In the Bayesian model averaging \citep[BMA;][]{rgbp} approach the forecast distribution is given by a weighted mixture of parametric densities, each of which depends on a single ensemble member with mixture weights being determined by the performance of the ensemble members in the training period. Within this article we build on the conceptually simpler ensemble model output statistics (EMOS) approach proposed by \citet{grwg}, where the conditional distribution of the weather variable of interest given the ensemble predictions is modeled by a single parametric family. The parameters of the forecast distribution are connected to the ensemble forecast through suitable link functions. For example, the original EMOS approach models temperature with a Gaussian predictive distribution where the mean is an affine function of the ensemble  member forecasts, and the variance is an affine function of the ensemble variance. 

Over the last years the EMOS approach has been extended to other weather variables such as wind speed \citep{tg, lt, bl15, schmoe}, precipitation \citep{sch, schham, bn}, and total cloud cover \citep{tcc}. The success of statistical post-processing relies on finding appropriate parametric families for the weather variable of interest. However, the choice of a suitable parametric model is a non-trivial task and often a multitude of competing models is available. The relative performance of these models usually varies for different data sets and applications. 

Regime-switching combination models proposed by \citet{lt} partly alleviate the limited flexibility of single parametric family models by selecting one of several candidate models based on covariate information. However, the applicability of this approach is subject to the availability of suitable covariates. For some weather variables, full mixture EMOS models can be formulated where the parameters and weights of a mixture of two forecast distributions are jointly estimated \citep{bl16}. However, such approaches are limited to specific weather variables, and the estimation is computationally demanding.

In this article we investigate the feasibility of another, more generally applicable route towards improving the forecast performance that has recently received some interest, and has for example been suggested in \citet{YangEtAl2017}. Motivated by promising results of \citet{MoellerGross2016} and \citet{BassettiEtAl2015}, we study whether combining predictive distributions from individual post-processing models is able to significantly improve the forecast performance. In a first step, individual EMOS models based on single parametric distributions are estimated. In a second step the forecast distributions are combined utilizing state of the art forecast combination techniques such as the (spread-adjusted) linear pool, the beta-transformed linear pool \citep{GneitingRanjan2013}, and a recently proposed Bayesian, essentially non-parametric calibration approach \citep{BassettiEtAl2015}. Further, we propose a computationally efficient 'plug-in' approach to determining combination weights in the linear pool that is specific to post-processing applications. 

The remainder of this article is organized as follows. Section \ref{sec:sec2} contains a description of the ensembles and the observation data. In Section \ref{sec:sec3}, the EMOS method is reviewed, and the individual EMOS models for wind speed and precipitation are introduced. Thereafter, Section \ref{sec:sec-new} provides a description of the forecast combination approaches and the application to post-processing. The various EMOS models and forecast combination approaches are compared in four case studies in Section \ref{sec:sec4}. The article concludes with a discussion in Section \ref{sec:sec5}.

\section{Data}
  \label{sec:sec2}
We consider two different weather variables, wind speed and precipitation accumulation, and two distinct data sets of ensemble forecasts and corresponding validating observations for each weather quantity. The wind speed data sets are identical to data used in \citet{bl15,bl16}, whereas the precipitation data coincide with those studied in \citet{bn}. For detailed descriptions of the ensemble forecasts and corresponding observations we refer to these articles and references therein. 

Ensemble members that are generated with the help of random perturbations of initial conditions are statistically indistinguishable, and are referred to as exchangeable. The notion of exchangeability of ensemble members is important for the formulation of post-processing models, see Section \ref{sec:sec3} for details.

\subsection{University of Washington mesoscale ensemble}
  \label{subs2.1}

The UWME covers the Pacific Northwest region of North America with a horizontal resolution of 12~km and consists of eight members generated from different runs of the fifth generation Pennsylvania State University--National Center for Atmospheric Research mesoscale model \citep{grell}. The initial and boundary conditions of the model runs are provided by different sources, the individual ensemble members can therefore be clearly distinguished and are considered to be non-exchangeable. The data set at hand contains 48~h ahead forecasts and corresponding validating observations for 10~m maximal wind speed (given in m/s) and 24~h precipitation accumulation (given in mm) for 152 stations in the Automated Surface Observing Network \citep{asos} in the U.S.~states of Washington, Oregon, Idaho, California and Nevada.

We focus on calendar year 2008 with additional forecasts and observations from the last months of 2007 used to allow for training periods of equal length for the model estimation. After removing days and locations with missing predictions and/or observations, stations where data are available only on very few days are also removed resulting in 101 stations with 27\,481 individual forecast cases for wind speed and 75 stations with 20\,448 individual forecast cases for precipitation.

\subsection{ALADIN-HUNEPS ensemble}
  \label{subs2.2}

The ALADIN-HUNEPS system covers large parts of continental Europe on an 8 km grid. It is obtained with dynamical downscaling of the global ARPEGE based PEARP system of M\'et\'eo France \citep{horanyi,dljbac} and consists of 11 ensemble members, 10 of which are exchangeable forecasts from perturbed initial conditions, and one of which is a control member from the unperturbed analysis.

We use ensembles of 42 h ahead forecasts of 10 m instantaneous wind speed (in m/s) and 24 h precipitation accumulation (in mm) issued for 10 major cities in Hungary together with the corresponding validation observations. Wind speed data are available for a one-year period from 1 April 2012 to 31 March 2013, and precipitation data are available between 1 October 2010 and 25 March 2011. Days with missing forecasts and/or observations are excluded from the analysis for both wind speed (6 days) and precipitation (2 days). 

\section{Ensemble model output statistics}
  \label{sec:sec3}

Successful statistical post-processing of ensemble forecasts relies on finding and estimating appropriate parametric models for the conditional distribution of the weather variable of interest given the ensemble predictions. In case of the EMOS approach, the forecast distribution is given by a single parametric law
with parameters depending on the ensemble forecast. While temperature can be modeled by a normal distribution 
\citep{grwg}, the choice of a suitable parametric family is much less straightforward for weather variables such as wind speed or precipitation. A multitude of post-processing approaches and modeling strategies has been proposed over the last years. In the following short review, we focus on EMOS models for wind speed and precipitation, and subsequently investigate methods to combine forecast distributions from different models.

\subsection{EMOS models for wind speed}
  \label{subs3.1}

Non-negative weather variables such as wind speed require skewed predictive distributions with non-negative support like Weibull \citep{jhmg} or gamma distributions \citep{gtpf}. Recently developed EMOS approaches utilize truncated normal \citep[TN;][]{tg}, gamma \citep{schmoe}, generalized extreme value \citep[GEV;][]{lt} and log-normal \citep[LN;][]{bl15} distributions to model the conditional distribution of wind speed given the ensemble predictions. Here, we focus on the truncated normal and log-normal model.

\subsubsection{Truncated normal EMOS model}

We denote by \ ${\mathcal N}_0\big(\mu,\sigma^2\big)$ \ the TN
distribution with location \ $\mu$, \ scale \ $\sigma>0$, \ and 
cut-off at zero with probability density function (PDF) 
\begin{equation*}
g(x\vert\, \mu,\sigma):=\frac{\frac
  1{\sigma}\varphi\big((x-\mu)/\sigma\big)}{\Phi\big(\mu/\sigma\big)
}, \quad \text{if $x\geq 0$,} \qquad \text{and} \qquad g(x\vert\, \mu,\sigma):=0,
\quad \text{otherwise,}
\end{equation*}
where \ $\varphi$ \ and \ $\Phi$ \ are the PDF and the cumulative distribution function (CDF) of the standard normal distribution, respectively. 
The predictive distribution of the EMOS model proposed by \citet{tg} is 
\begin{equation}
   \label{eq:eq3.1}
  {\mathcal N}_0\big(a_0+a_1f_1+ \cdots +a_Kf_K,b_0+b_1S^2\big) \qquad \text{with} \qquad S^2:=\frac 1{K\!-\!1}\sum_{k=1}^K\big (f_k-\overline f\big)^2,
\end{equation}
where  \ $f_1,f_2,\ldots ,f_K$ \ denote the ensemble of
distinguishable forecasts
of wind speed for a given location and time, and \ $\overline f$ \ is the ensemble mean. Location parameters \ $a_0, a_1, \ldots ,a_K \in\mathbb{R}$ \ and scale parameters
\ $b_0  \in\mathbb{R}, \ b_1 \geq0$ \  of model \eqref{eq:eq3.1} can be estimated from the training data, consisting of ensemble members and verifying observations from the preceding \ $n$ \ days, by optimizing an appropriate verification score, see Section \ref{subs3.3}.

However, most operational EPSs generate forecasts using random perturbations of the initial conditions resulting in statistically indistinguishable ensemble members which are referred to as exchangeable. Examples include the 51-member ECMWF ensemble, as well as sub-ensembles of forecasts from single models that form groups of exchangeable members within
multi-model EPSs such as the THORPEX Interactive Grand Global Ensemble \citep{tigge16} or the GLAMEPS ensemble \citep{iversen11}. To account for the generation of the forecasts, ensemble members within a given group of exchangeable members should share the same coefficients in the post-processing model \citep{frg2010,gneiting14}. 

To formalize this notion, a generalized formulation of model \eqref{eq:eq3.1} for the case of  \ $M$ \ ensemble members divided into  \ $K$ \  groups, where the \ $k$th \ group contains \ $M_k\geq 1$ \ exchangeable ensemble members ($\sum_{k=1}^KM_k=M$) introduced in \citet{bl15} is given by
\begin{equation*}
  {\mathcal N}_0\bigg(a_0+a_1\sum_{\ell_1=1}^{M_1}f_{1,\ell_1}+ \cdots
  +a_K\sum_{\ell_K=1}^{M_K} f_{K,\ell_K},b_0+b_1S^2\bigg). 
\end{equation*}
Analogous concepts apply to all EMOS models discussed in the subsequent sections.

\subsubsection{Log-normal EMOS model}

As an alternative to the TN EMOS model, \citet{bl15} introduce an EMOS approach based on log-normal forecast distributions where the mean \ $m$ \ and  variance \ $v$ \ of the predictive distribution are linked to the ensemble members as 
\begin{equation}
\label{eq:eq3.2}
m=\alpha_0+\alpha_1f_1+ \cdots +\alpha_Kf_K \qquad\text{and}\qquad
v=\beta_0+\beta_1 S^2.
\end{equation}
These quantities uniquely determine the location \ $\mu$ \ and shape \ $\sigma>0$ \ of the underlying LN distribution \ $\mathcal{LN}\big(\mu,\sigma\big)$ \ with PDF
\begin{equation*}
h(x\vert\, \mu,\sigma):=\frac
  1{x\sigma}\varphi\big((\log x-\mu)/\sigma\big), \quad \text{if $x\geq 0$,}
  \qquad \text{and} \qquad h(x\vert\, \mu,\sigma):=0, 
\quad \text{otherwise,}
\end{equation*}
via transformations
\begin{equation*}
\mu =\log \bigg(\frac {m^2}{\sqrt{v+m^2}}\bigg) \qquad \text{and}
\qquad \sigma=\sqrt{\log \Big (1+ \frac v{m^2}\Big)}.
\end{equation*}
Similar to the TN EMOS model, estimates of parameters \ $\alpha_0, \alpha_1, \ldots, \alpha_K \in\mathbb{R}$ \ and \ $\beta_0  \in\mathbb{R}, \ \beta_1 \geq0$  are obtained by optimizing the mean of an appropriate verification score over all forecast cases in the training data.

\subsubsection{Combination and mixture models}

The TN and LN models described above model the conditional distribution of wind speed given the ensemble predictions with a single parametric forecast distribution. This approach relies on the choice of a suitable parametric family, and limits  the flexibility of the model. For instance, it can be demonstrated that the heavier tails of the LN model are more appropriate for modeling higher wind speeds in the right tail of the distribution, whereas the TN model is more appropriate for the bulk of the distribution, see \citet{bl16} for details.

Therefore, different combination and mixture models have been proposed in the literature. In the regime-switching combination approach \citep{lt,bl15} one of the candidate models is selected based on covariate information with suitably adapted parameter estimation procedures. For example, a TN model can be used if the median ensemble forecast is below a threshold $\ \eta,\ $ and an LN model is used in case of median ensemble forecasts exceeding this threshold. Such combination models have been demonstrated to improve the predictive performance compared to the individual models, however, they require the choice of a suitable covariate, and the threshold parameter $\ \eta\ $ has to be determined by repeating the model estimation over a grid of potential values, thereby limiting the flexibility and increasing the computational cost of such approaches.

In order to flexibly combine the advantages of lighter and heavier-tailed distributions and to avoid these problems in the process, \citet{bl16} propose a mixture model of the form
\begin{equation}
   \label{eq:eq3.3}
\psi(x\vert\, \mu_{TN},\sigma_{TN};\mu_{LN},\sigma_{LN};\omega):=\omega g(x\vert\, \mu_{TN},\sigma_{TN})+(1-\omega)h(x\vert\, \mu_{LN},\sigma_{LN}),
\end{equation}
where the parameters of the component distributions  $\ g\ $ and $\ h\ $ depend on the ensemble forecasts as specified in \eqref{eq:eq3.1} and \eqref{eq:eq3.2}. The EMOS coefficients and the weight \ $\omega \ \in [0,1]$ \ of the mixture model \eqref{eq:eq3.3} are estimated jointly using optimum score approaches. This mixture model approach results in significantly improved calibration \citep{bl16}, however, it is computationally very demanding and hinders using standard optimum score estimation based on the continuous ranked probability score due to the lack of an analytic expression of the objective function, see Section \ref{subs3.3} for details. Similar mixture models where the different component distributions focus on specific regions of interest such as the bulk and the tail above a threshold value have been tested, but models based on truncated normal and generalized Pareto distributions result in worse predictive performance, see \citet{bl16}.

In contrast to the joint estimation of all parameters in \eqref{eq:eq3.3}, the forecast combination approaches 
introduced in Section \ref{sec:sec-new} are two-step procedures where in a first step, EMOS models based on a single parametric family are estimated, and in a second step, these models are combined as a weighted mixture by estimating an appropriate weight. In Section~\ref{sec:sec4} the full mixture model \eqref{eq:eq3.3}
is used as a benchmark, whereas the regime-switching combination approach will not be considered any further.

\subsection{EMOS models for precipitation}
  \label{subs3.2}

The discrete-continuous nature of precipitation accumulation requires a non-negative predictive distribution assigning positive mass to the event of zero precipitation. A popular choice is to consider a continuous distribution that can take both positive and negative values and left censor it at zero \citep{sch,schham,bn}, which thereby assigns the mass of negative values to zero precipitation accumulation.

\subsubsection{Censored and shifted gamma EMOS model}

Let \ $G(\cdot|\, \kappa,\theta)$ \ denote the CDF of the \ $\Gamma (\kappa,\theta)$ \ distribution with shape \ $\kappa>0$ \ and scale \ $\theta>0$ \ and let \ $\delta>0$. \ Then the CDF of the shifted gamma distribution left censored at zero (CSG) \ $\Gamma_0(\kappa,\theta,\delta)$ \ with shape \ $\kappa$, \ scale \ $\theta$ \ and shift \ $\delta$ \ is given by
\begin{equation}
  \label{eq:eq3.4}
 G_0(x|\,\kappa,\theta,\delta):=
   G(x+\delta |\,\kappa,\theta), \ \ \text{if \ $x\geq 0$,}
\qquad \text{and} \qquad G_0(x|\,\kappa,\theta,\delta):=0, 
\ \  \text{otherwise,}
\end{equation} 
that is, mass \ $G(\delta|\,\kappa,\theta)$ \ is assigned to the origin. 
In the CSG EMOS approach of \citet{bn} the mean \ $m=\kappa\theta $ \ and variance \ $\sigma^2=\kappa\theta ^2$ \ of the uncensored gamma distribution \ $\Gamma (\kappa,\theta)$ \ are affine functions of the ensemble and ensemble mean, respectively, that is
\begin{equation}
  \label{eq:eq3.5}
m = a_0+a_1f_1+ \cdots +a_Kf_K \qquad \text{and} \qquad  \sigma^2 = b_0+b_1\overline f.
\end{equation}

\subsubsection{Censored generalized extreme value EMOS model}

The CDF of a GEV distribution \ $\mathcal{GEV}\big(\mu,\sigma,\xi\big)$ \ with location \ $\mu $, \ scale \ $\sigma>0$ \ and shape \ $\xi$ \ equals
\begin{equation*}
H(x|\,\mu,\sigma,\xi):=\begin{cases}
\exp\Big(-\big[1+\xi(\frac{x-\mu}{\sigma})\big]^{-1/\xi}\Big), & \
\xi\ne 0; \\
\exp\Big(-\exp\big(-\frac{x-\mu}{\sigma}\big)\Big), & \
\xi= 0,
\end{cases} \qquad \quad \text{if $1+\xi(x-\mu)/\sigma> 0$,}
\end{equation*}  
and zero otherwise, which for \ $-0.278<\xi<1$ \ has a positive skewness and an existing mean 
\begin{equation*}
m=\begin{cases}
\mu+\sigma \frac {\Gamma(1-\xi)-1}{\xi} , & \
\xi\ne 0; \\
\mu +\sigma \gamma, & \
\xi= 0, 
\end{cases}
\end{equation*}
where \ $\gamma $ \ denotes the Euler-Mascheroni constant.

The EMOS model for precipitation accumulation proposed by \citet{sch} is based on a censored  GEV distribution  \ $\mathcal{GEV}_0\big(\mu,\sigma,\xi\big)$ \ with CDF
\begin{equation}
  \label{eq:eq3.6}
 H_0(x|\,\mu,\sigma,\xi)=
   H(x|\,\mu,\sigma,\xi), \ \ \text{if \ $x\geq 0$,}
\qquad \text{and} \qquad H_0(x|\,\mu,\sigma,\xi):=0, 
\ \ \text{otherwise,}
\end{equation} 
where
\begin{equation}
    \label{eq:eq3.7}
m = \alpha_0+\alpha_1f_1+ \cdots +\alpha_Kf_K + \nu p_0 \qquad \text{and} \qquad  \sigma = \beta_0+\beta_1\md(f), 
\end{equation} 
with
\begin{equation*}
p_0:=\frac 1K \sum_{k=1}^K{\mathbbm{1}}_{\{f_k=0\}} \qquad \text{and} \qquad \md(f):=\frac 1{K^2}\sum_{k,\ell=1}^K\big |f_k-f_{\ell}\big|,
\end{equation*}
where \ $\mathbbm 1_A$ \ denotes the indicator function of the set \ $A$. 

\subsubsection{Mixture models}

Similar to wind speed, general mixture models with CSG and GEV component distributions of the form
\begin{equation}
   \label{eq:eq3.8}
\varrho(x\vert\, \kappa,\theta,\delta;\mu,\sigma,\xi;\omega):=\omega g_0(x|\,\kappa_{CSG},\theta_{CSG},\delta_{CSG}) +(1-\omega)h_0(x|\,\mu_{GEV},\sigma_{GEV},\xi_{GEV}),
\end{equation}
can be formulated, where  \ $g_0(\cdot|\,\kappa,\theta,\delta)$ \ and \ $h_0(\cdot |\,\mu,\sigma,\xi)$ \ denote the generalized PDFs of the CSG and censored GEV distributions, respectively, and the dependence of parameters \ $\kappa_{CSG},\theta_{CSG}$ \ and \ $\mu_{GEV},\sigma_{GEV}$ \ on the ensemble is given by \eqref{eq:eq3.5} and \eqref{eq:eq3.7}.  

However, joint optimum score estimation of the parameters is more involved than in the case of the TN-LN mixture model \eqref{eq:eq3.3} for wind speed due to the larger number of parameters and the discrete-continuous nature of the forecast distribution \ $\varrho.\ $ As initial tests with the precipitation data sets introduced in Section \ref{sec:sec2} indicated problematic behavior of the numerical optimization algorithms potentially caused by the non-smooth dependence of the objective functions on the parameters, we do not pursue this approach any further and only consider the forecast combination approaches introduced in Section \ref{sec:sec-new}. Compared to jointly estimating all parameters, these methods separate the estimation into two steps and thereby result in more stable optimization problems.

\subsection{Forecast evaluation and parameter estimation}
  \label{subs3.3}

In probabilistic forecasting the general aim is to maximize the sharpness of the predictive distribution subject to calibration \citep{gbr}. Calibration refers to the statistical consistency between the forecast and the observation, and given that the predictive distribution is calibrated, it should be as concentrated (or sharp) as possible. Calibration and sharpness can be assessed simultaneously with the help of proper scoring rules. 

Proper scoring rules are loss functions that assign a numerical value to pairs of forecasts and observations. In the atmospheric sciences the most popular scoring rules are the continuous ranked probability score \citep[CRPS;][]{MathesonWinkler1976,grjasa} and the logarithmic score \citep[LogS;][]{Good1952}. Given a predictive CDF \ $F(y)$ \ and an observation \ $x$, \ the CRPS is defined as
\begin{align}
  \label{eq:eq3.9}
\crps\big(F,x\big):=&\int_{-\infty}^{\infty}\big (F(y)-{\mathbbm 
  1}_{\{y \geq x\}}\big )^2\,{\mathrm d}y \\=&\,\int_{-\infty}^x F^2(y){\mathrm d} y +\int_x^{\infty}\big (1-F(y)\big )^2{\mathrm d}y \nonumber  \\
   = & \ {\mathsf E}|X-x|-\frac 12 {\mathsf E}|X-X'|, \nonumber
\end{align}
where \ $X$ \ and \ $X'$ \ are independent random variables with CDF \
$F$ \ and finite first moment. The last representation in \eqref{eq:eq3.9} implies that the CRPS can be expressed in the same unit as the observation.
The logarithmic score is the negative logarithm of the predictive density \ $f(y)$ \ evaluated at the verifying observation, i.e.,
\begin{equation*}
 \logs(F,x) := - \log(f(x)).
\end{equation*}
Both CRPS and LogS are proper scoring rules \citep{grjasa} which are negatively oriented, that is, smaller scores indicate better forecasts.

Proper scoring rules provide valuable tools for the estimation of model parameters. Following the general optimum score estimation approach of \citet{grjasa}, the parameters of a predictive distribution can be determined by optimizing the average value of a proper scoring rule as a function of the parameters over a suitably chosen training set. Optimum score estimation based on minimizing the LogS then corresponds to classical maximum likelihood (ML) estimation. If closed form expressions of the integral in \eqref{eq:eq3.9} are available, minimum CRPS estimation, i.e.~optimum score estimation based on minimizing the mean CRPS, often provides a valuable, more robust alternative to ML estimation. 

Analytic expressions of the CRPS are available for all individual EMOS models for wind speed and precipitation introduced in Sections \ref{subs3.1} and \ref{subs3.2}, thereby allowing for efficient parameter estimation procedures by minimizing the mean CRPS over the forecast cases in the training periods. The closed form solutions are provided in the corresponding articles \citep{tg,bl15,sch,schham}. Implementations for the statistical programming language \texttt{R} \citep{R} are for example available in the \texttt{scoringRules} package \citep{scoringRules}. The parameter estimation for the EMOS models is performed using the Broyden-Fletcher-Goldfarb-Shanno (BFGS) algorithm \citep[Section 10.9]{press} implemented in the \texttt{optim} function in \texttt{R}. In the case of precipitation we use a constrained version of the BFGS algorithm to ensure positivity of the EMOS coefficients. See Section \ref{sec:sec4} for details on the selection of the training sets over which the parameters are estimated.
 
By contrast, the CRPS is not available in closed form for the mixture models \eqref{eq:eq3.3} and \eqref{eq:eq3.8}, or any of the forecast combination models introduced in Section \ref{sec:sec-new}. Therefore, each step of the optimization procedure requires numerical integration resulting in high computational costs. In case of the mixture model  \eqref{eq:eq3.3} for wind speed, we instead use ML estimation of the parameters. The forecast combination approaches introduced below partly alleviate this issue by separating the parameter estimation into two steps rather than estimating all parameters jointly.

\section{Forecast combination methods and application to statistical post-processing}
  \label{sec:sec-new}

We now describe state of the art methods for combining predictive distributions, which we will employ in a post-processing context. The combination approaches constitute two-step methods. The first step is given by the estimation of component models in the form of EMOS models based on suitable single parametric families. In a second step, the component models are combined by estimating the mixture weight and possibly more combination parameters. Compared to the previously discussed mixture model \eqref{eq:eq3.3}, the two-step approaches reduce the dimensionality of the optimization problem. Further, the combination approaches can be flexibly applied to any weather variable of interest given that suitable component models are available, the model formulation is thus given in a general form. 

Let \ $G(x|\,\boldsymbol f; \nu)$ \ and \ $H(x|\,\boldsymbol f; \theta)$ \ be predictive CDFs belonging to two different families of distributions depending on the ensemble \ $\boldsymbol f$ \ via parameter vectors \ $\nu$ \ and \ $\theta$, \ respectively. The EMOS models introduced in Sections \ref{subs3.1} and \ref{eq:eq3.2} are later used as component distributions for wind speed and precipitation, respectively.

\subsection{Linear pool and spread-adjusted linear pool}
\label{sec:lpslp}

We start by introducing the linear and spread-adjusted linear pool of forecast distributions which have been applied to post-processing ensemble forecasts by \citet{MoellerGross2016}. The classical linear pool (LP) employs a mixture model with a predictive CDF of the general form
\begin{equation}
\label{eq:Flp}
F^\textnormal{LP} (x|\, \boldsymbol f; \nu,\theta,\omega):=\omega \,G(x|\,\boldsymbol f; \nu) + (1-\omega)H(x|\,\boldsymbol f; \theta), \qquad \omega \in [0,1].
\end{equation}

As demonstrated by \citet{GneitingRanjan2013}, linear pooling of predictive distributions increases the dispersion of the forecasts. They propose spread-adjusted and beta-transformed linear pooling approaches that allow to correct for this deficiency. The spread-adjusted linear pool (SLP) results in a predictive distribution
 \begin{equation}
 \label{eq:eq3.12}
 F^\textnormal{SLP} (x|\, \boldsymbol f; \nu,\theta,\omega):=\omega \,G\left(\frac{x}{c}\big|\,\boldsymbol f; \nu \right) + (1-\omega)H\left( \frac{x}{c}\big|\,\boldsymbol f; \theta\right), \qquad \omega \in [0,1]
 \end{equation}
with spread adjustment parameter \ $c > 0.\ $ The linear pool is obtained for \ $c = 1.\ $

As noted by \citet{GneitingRanjan2013}, the forecasts of the Bayesian model averaging approach of \citet{rgbp} take a similar functional form, but differ in that the combination parameters and the parameters of the individual component distributions are estimated simultaneously. Further, the forecast distributions of the mixture components in the BMA approach depend on a single ensemble member only, whereas the EMOS predictive distributions used here depend on the entire ensemble through suitable link functions with coefficients $\ \nu,\theta$.

The weight $\ \omega \in [0,1]\ $ and the spread adjustment parameter $\ c > 0\ $ have to be estimated from past forecast cases. Note that these need to be training samples where post-processed forecast distributions are available. \citet{MoellerGross2016} suggest to choose sets of candidate parameter values, e.g.$\ \omega\in\{0, 0.05, \dots, 0.95, 1\}\ $ and $\ c\in\{0.7, 0.75, \dots, 1.25, 1.3\},\ $ and to apply the combination formulas \eqref{eq:Flp} and \eqref{eq:eq3.12} for all possible parameter combinations in order to select those parameter values corresponding to the lowest mean CRPS in the training sample. The CRPS of the forecast distributions is thereby computed using numerical integration.  However, as tests indicated improvements in the predictive performance and lower computational costs, we instead determine the optimal parameter values by numerical optimization with the CRPS as a target functional. To allow for a direct comparison with the component models, we use forecast-observation pairs from the same training sets that were used to estimate the EMOS coefficients. The use of training sets expanding over time or separately estimating the combination parameters for all stations could potentially result in further improvements. Various alternative approaches to estimate the combination weight in case of the linear pool have been proposed in the literature, including approaches based on other scoring rules such as the LogS \citep{hm07} or weighted scoring rules \citep{OpschoorEtAl2017}, as well as Bayesian approaches \citep{bcrvd,dnhs}.


\subsubsection{A plug-in variant of the linear pool for post-processing applications}

In the following we propose a simple plug-in variant to determine the weight parameter in the linear pool that only requires a single numerical integration step rather than repeated numerical integration during the optimization procedure. In tables and figures of Section \ref{sec:sec4}, this approach is abbreviated by LP-PI.

Let \ $G(x|\,\boldsymbol f; \nu)$ \ and \ $H(x|\,\boldsymbol f; \theta)$ \ be predictive CDFs as above, and let \ $(x_k,\boldsymbol f_k), \ k=1,2,\ldots ,n,$ \ denote the pairs of verifying observations and ensemble forecasts in the training data. The basic idea of the proposed plug-in variant of the linear pool is to utilize the current EMOS parameters estimated for day \ $n+\tau$ \ where \ $\tau$ \ is the forecast horizon to compute the parameters of the corresponding component distributions over the entire training sample. In contrast to utilizing the respective EMOS coefficient vectors \ $(\nu_k)$ \ and \ $(\theta_k)$ \ for \ $k=1,2,\dots,n$, \ this reduces the number of required numerical integrations at the cost of not using the specific parameter values estimated for those days in the training period. Since no spread adjustment is applied, this approach shares the deficiencies of the standard linear pool described above.

As before, consider a linear pool mixture model with predictive CDF 
\begin{equation}
\label{eq:eq3.10}
F^{\text{LP}} (x|\, \boldsymbol f; \nu,\theta,\omega):=\omega \,G(x|\,\boldsymbol f; \nu) + (1-\omega)H(x|\,\boldsymbol f; \theta), \qquad \omega \in [0,1].
\end{equation}
Short calculation based on the integral representation in the second line of \eqref{eq:eq3.9} shows
\begin{align*}
\crps&\,\big(F^{\text{LP}}(\cdot |\, \boldsymbol f; \nu,\theta,\omega),x\big)=\omega ^2 \crps \big(G(\cdot|\,\boldsymbol f; \nu),x\big) + (1-\omega )^2 \crps\big(H(\cdot|\,\boldsymbol f; \theta),x\big) \\
&+2\omega(1-\omega)\left[ \int_{-\infty}^x G(y|\,\boldsymbol f; \nu) H(y|\,\boldsymbol f; \theta){\mathrm d}y +\int_x^{\infty} \big(1-G(y|\,\boldsymbol f; \nu)\big)\big(1- H(y|\,\boldsymbol f; \theta)\big){\mathrm d}y \right].
\end{align*}

Let  \ $\nu_{\circ}$ \ and \ $\theta_{\circ}$ \ denote the optimal parameters of the individual models in the training set estimated in the first step for day $n+\tau$, that is
\begin{equation*}
\nu_{\circ}:=\arg\min_\nu \overline {\crps (G, \nu)} \quad \text{and} \quad \theta_{\circ}:=\arg\min_\theta \overline {\crps (H, \theta)},
\end{equation*}
where
\begin{equation*}
\overline {\crps (G, \nu)}\!:=\!\frac 1n\!\sum_{k=1}^n \!\crps \big(G(\cdot|\,\boldsymbol f_k; \nu),x_k\big), \qquad
\overline {\crps (H, \theta)}\!:=\!\frac 1n\!\sum_{k=1}^n \!\crps \big(H(\cdot|\,\boldsymbol f_k; \theta),x_k\big).
\end{equation*}

We propose to use \ $\nu_{\circ}$ \ and \ $\theta_{\circ}$ \ as parameters of the mixture model \eqref{eq:eq3.10} and then, in the modified second step, to optimize
\begin{equation*}
\overline{\crps (F^{\text{LP}},\omega)}:=\frac 1n \sum_{i=1}^n\crps\big(F^{\text{LP}}(\cdot |\, \boldsymbol f_k; \nu_{\circ},\theta_{\circ},\omega),x_k\big)
\end{equation*}
as a function of \ $\omega.$ \ The minimum point of \ $\overline{\crps (F^{\text{LP}},\omega)}$ \ is 
\begin{equation*}
\omega_{\circ}^*=\frac{\overline {\crps (H, \theta_{\circ})}-\overline{{\mathcal M}(G,H,\nu_{\circ},\theta_{\circ})}}{\overline {\crps (G, \nu_{\circ})}+\overline {\crps (H, \theta_{\circ})}-2\overline{{\mathcal M}(G,H,\nu_{\circ},\theta_{\circ})}},
\end{equation*}
where
\begin{equation*}
\overline{{\mathcal M}(G,H,\nu,\theta)}\!:=\!\frac 1n \sum_{k=1}^n\! \left[\int_{-\infty}^{x_k} \!\!\!\!\!G(y|\,\boldsymbol f_k; \nu) H(y|\,\boldsymbol f_k; \theta){\mathrm d}y \!+\!\!\int_{x_k}^{\infty} \!\!\!\!\!\big(1\!-\!G(y|\,\boldsymbol f_k; \nu)\big)\big(1\!-\! H(y|\,\boldsymbol f_k; \theta)\big){\mathrm d}y\right]\!,
\end{equation*}
and short calculation shows
\begin{align*}
\omega_{\circ}^*=&\,\frac{\sum_{k=1}^n\int_{-\infty}^{\infty} H(y|\,\boldsymbol f_k; \theta_{\circ})\big(H(y|\,\boldsymbol f_k; \theta_{\circ})- G(y|\,\boldsymbol f_k; \nu_{\circ})\big){\mathrm d}y}{\sum_{k=1}^n\int_{-\infty}^{\infty} \big(H(y|\,\boldsymbol f_k; \theta_{\circ})- G(y|\,\boldsymbol f_k; \nu_{\circ})\big)^2{\mathrm d}y}\\[2mm]
&-\frac{\sum_{k=1}^n\int_{x_k}^{\infty} \big(H(y|\,\boldsymbol f_k; \theta_{\circ})- G(y|\,\boldsymbol f_k; \nu_{\circ})\big){\mathrm d}y }{\sum_{k=1}^n\int_{-\infty}^{\infty} \big(H(y|\,\boldsymbol f_k; \theta_{\circ})- G(y|\,\boldsymbol f_k; \nu_{\circ})\big)^2{\mathrm d}y}.
\end{align*}

Now, as \ $\omega_{\circ}^*$ \ might fall outside the unit interval \ $[0,1],$ \ we use
\begin{equation*}
\omega_{\circ}:=\min \big\{\max\, \{0,\omega_{\circ}^*\},1\big\}
\end{equation*}
as our final estimate of the weight. Finally, one can easily show that within the training sample
\begin{equation}
\label{eq:eq3.11}
\overline{\crps (F^{\text{LP}},\omega_{\circ})} \leq \min \big\{\overline {\crps (G, \nu_{\circ})},\overline {\crps (H, \theta_{\circ})}\big\}, 
\end{equation}
so the mean CRPS of the mixture model \eqref{eq:eq3.10} with parameters \ $(\nu_{\circ},\theta_{\circ},\omega_{\circ})$ \ cannot exceed the optimal mean CRPS values of the components. Obviously, \eqref{eq:eq3.11} gives no guarantee that for a new out-of-sample pair \ $(\widetilde x,\widetilde{\boldsymbol f})$ \ the 
CRPS of the mixture \ $\crps\big(F^{\text{LP}}(\cdot |\, \widetilde{\boldsymbol f}; \nu_{\circ},\theta_{\circ},\omega_{\circ}),\widetilde x\big)$ \ does not exceed any of the corresponding individual CRPS values. 

The above method can be generalized to a convex combination of \ $r$ \ different parametric families. However, in this case the optimal weight vector is a coordinate-wise non-negative solution of a quadratic optimization problem with a single linear constraint, where the main diagonal of the corresponding  \ $r\times r$ \ symmetric matrix consists of the mean CRPS values of the component models, whereas the other entries, which are similar to \ $\overline{{\mathcal M}(G,H,\nu,\theta)}$, \ can be expressed via integrals.

\subsection{Beta-transformed linear pool} 

As an alternative to the spread-adjusted linear pool that allows for correcting for the lack of dispersion of the linear pool, \citet{GneitingRanjan2013} propose the beta-transformed linear pool (BLP) with predictive CDF
\begin{equation}
\label{eq:blp}
F^\textnormal{BLP} (x|\, \boldsymbol f; \nu,\theta,\omega):= B_{\alpha,\beta}\left(\omega \,G(x|\,\boldsymbol f; \nu) + (1-\omega)H(x|\,\boldsymbol f; \theta)\right), \qquad \omega \in [0,1].
\end{equation}
Here, \ $B_{\alpha,\beta}$ \ denotes the CDF of the beta distribution with parameters \ $\alpha > 0$ \ and \ $\beta > 0$. \ 

Similar to the linear and spread-adjusted linear pool, the combination parameters \ $\alpha,\beta,\omega$ \ have to be estimated from suitably chosen training data. We proceed as before and estimate the parameters by numerically optimizing the mean CRPS over the forecast cases that coincide with the training period used to determine the coefficients of the EMOS models. Note that the representation in the second line of equation \eqref{eq:eq3.9} with lower bound 0 is beneficial to avoid numerical issues when computing the CRPS using numerical integration, particularly for precipitation forecasts.

\subsection{Bayesian non-parametric combination approach}
\label{sec:bmc}

\citet{BassettiEtAl2015} recently proposed an extension of the BLP approach. Motivated by results on mixture distributions from theoretical statistics, they propose a forecast aggregation method based on a mixture of beta distributions. In case of a finite mixture with \ $L$ \ components, the resulting predictive CDF is 
\begin{equation}\label{eq:bmk}
F^{\textnormal{BM}_L}(x|\, \boldsymbol f; \nu,\theta,\omega):= \sum_{\ell=1}^{L}w_{\ell} B_{\alpha_{\ell},\beta_{\ell}}\left(\omega \,G(x|\,\boldsymbol f; \nu) + (1-\omega)H(x|\,\boldsymbol f; \theta)\right), \qquad \omega \in [0,1],
\end{equation} 
where \ $\alpha_{\ell} > 0,\ \beta_{\ell} > 0, \ w_{\ell}\geq 0, \  {\ell} = 1,2,\dots,L$, \ are the parameters of the beta mixture components. The BLP approach in \eqref{eq:blp} arises as a special case for \ $L = 1$. 

As the number \ $L$ \ of components is usually unknown, \citet{BassettiEtAl2015} propose a Bayesian inference approach that allows to treat \ $L$ \ as unbounded and random. This infinite beta mixture approach, referred to as BMC approach in the following, has CDF
\begin{equation}\label{eq:bmc}
F^{\textnormal{BM}_{\infty}}(x|\, \boldsymbol f; \nu,\theta,\omega):= \sum_{\ell=1}^{\infty}w_{\ell} B_{\alpha_{\ell},\beta_{\ell}}\left(\omega \,G(x|\,\boldsymbol f; \nu) + (1-\omega)H(x|\,\boldsymbol f; \theta)\right), \quad \omega \in [0,1].
\end{equation} 

Based on the slice sampling algorithm of \citet{walker} and \citet{kgw} for infinite mixtures, \citet{BassettiEtAl2015} give an algorithm that results in samples from mixture parameters \ $\alpha_{\ell},\beta_{\ell},w_{\ell}$ \ and \ $\omega$, \ allowing to generate draws from the predictive distribution \eqref{eq:bmc}. Note that the algorithm in fact deals with finite mixtures, however, the number of components may differ from draw to draw. In order to obtain an estimate of a verification score for a given location and time, we  average over the predictive CDFs obtained though iterations of the algorithm and compute the score for the mean CDF by discrete approximation of the first integral of \eqref{eq:eq3.9} over a dense grid. Note that this approach is in line with theoretical considerations on forecast evaluation based on simulation output discussed in \citet{kltg}.

In a case study based on wind speed data from a single observation station in \citet{BassettiEtAl2015}, the BMC method shows very promising results and substantially outperforms the linear pool. Here, we apply the BMC method to larger data sets of ensemble forecasts of wind speed and precipitation at multiple stations. Due to the point mass at zero precipitation in the forecast distributions, some minor adjustments of the sampling algorithm described in \citet{BassettiEtAl2015} are required. Specifically, in steps 4 and 5 of the Gibbs sampling algorithm described in Section S1.2 of the supplementary materials of \citet{BassettiEtAl2015}, in case of zero observed precipitation, random values from the intervals between zero and the corresponding probabilities of no precipitation are chosen as values of the component CSG and GEV predictive CDFs. We found that without this adjustment, the marginal predictive CDF generally underestimates precipitation accumulation substantially. The value of the resulting CDF at 0, i.e., $\ F^{\textnormal{BM}_{\infty}}(0|\, \boldsymbol f; \nu,\theta,\omega),\ $ is approximated by linear interpolation of the values at the first two grid points.

\section{Case studies}
  \label{sec:sec4}
  
Here, we report the results of four case studies for the wind speed and precipitation data sets introduced in Section \ref{sec:sec2}. 
Note that BMC results for each forecast case are based on the forecast distribution given by the mean of 50 predictive CDFs obtained from the post burn-in iterations of the sampling algorithm described in \citet{BassettiEtAl2015}.

\subsection{Wind speed} 
  \label{subs4.2}

\begin{figure}[p]
\begin{tabular}{cc}
\multicolumn{2}{c}{UWME} \\[0.5em]
(a) combination weight & (b) further parameters \\[0.5em]
\includegraphics[width=0.48\textwidth]{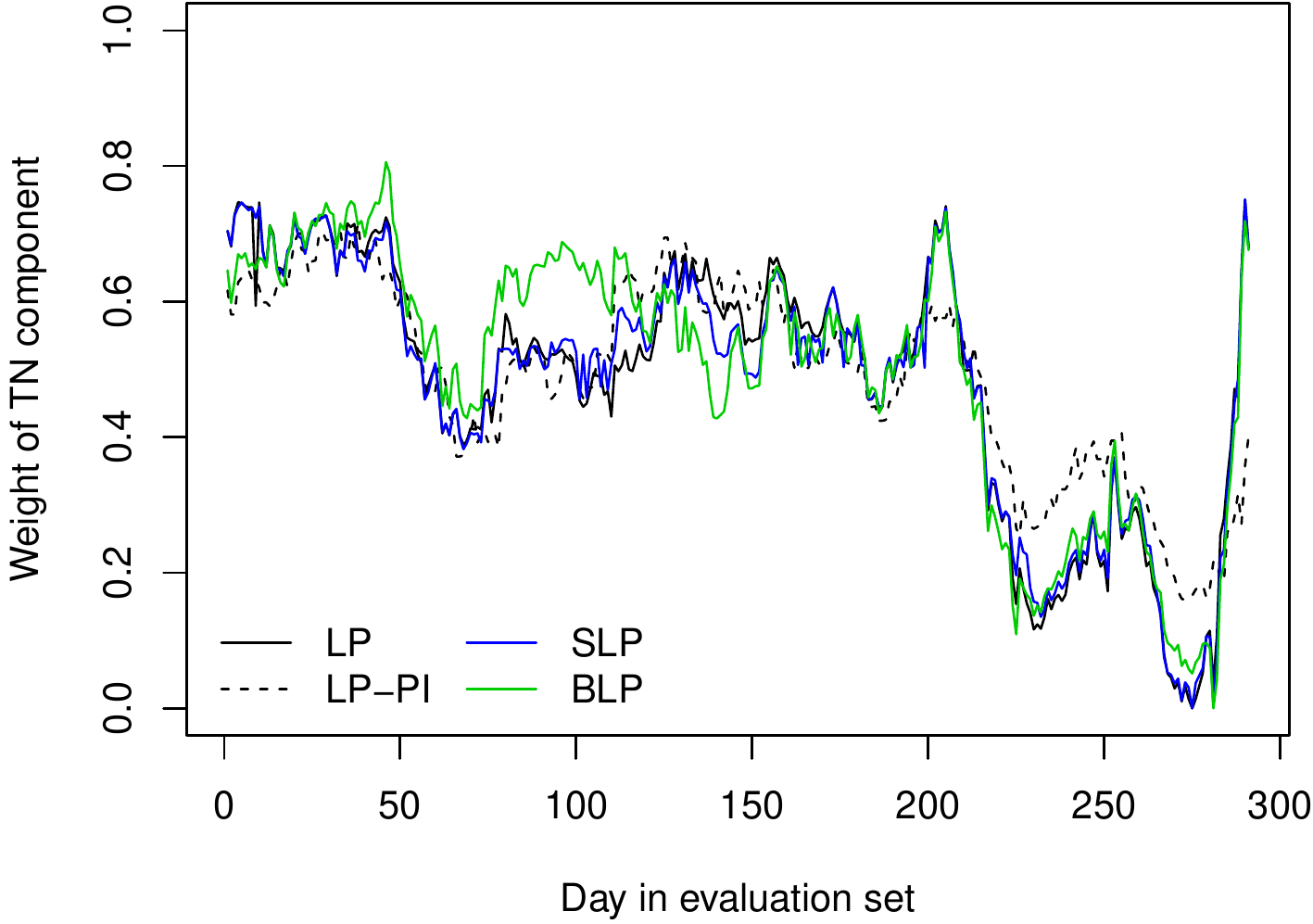} &
\includegraphics[width=0.48\textwidth]{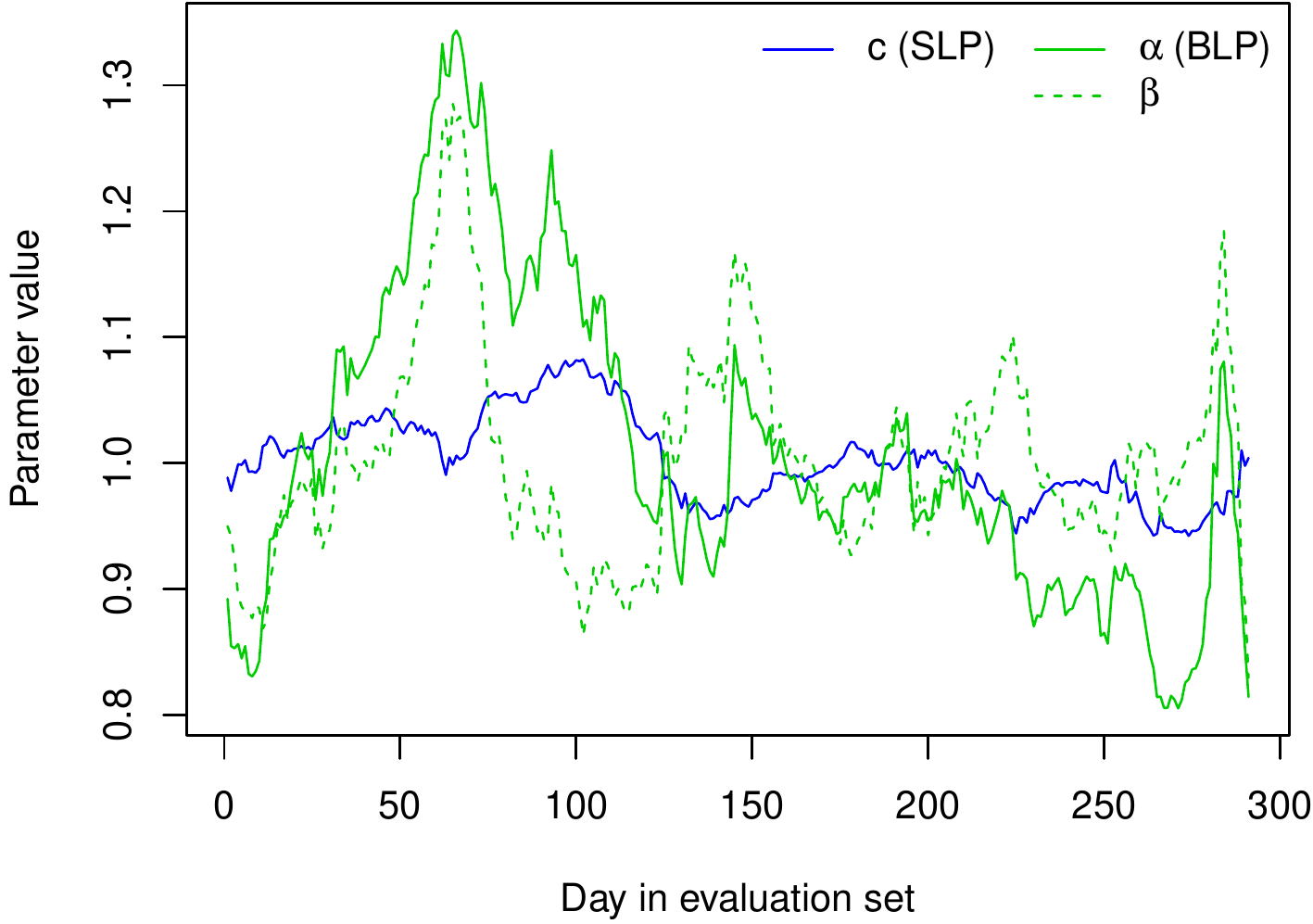} \\[1em]
\multicolumn{2}{c}{ALADIN-HUNEPS} \\[0.5em]
(c) combination weight & (d) further parameters \\[0.5em]
\includegraphics[width=0.48\textwidth]{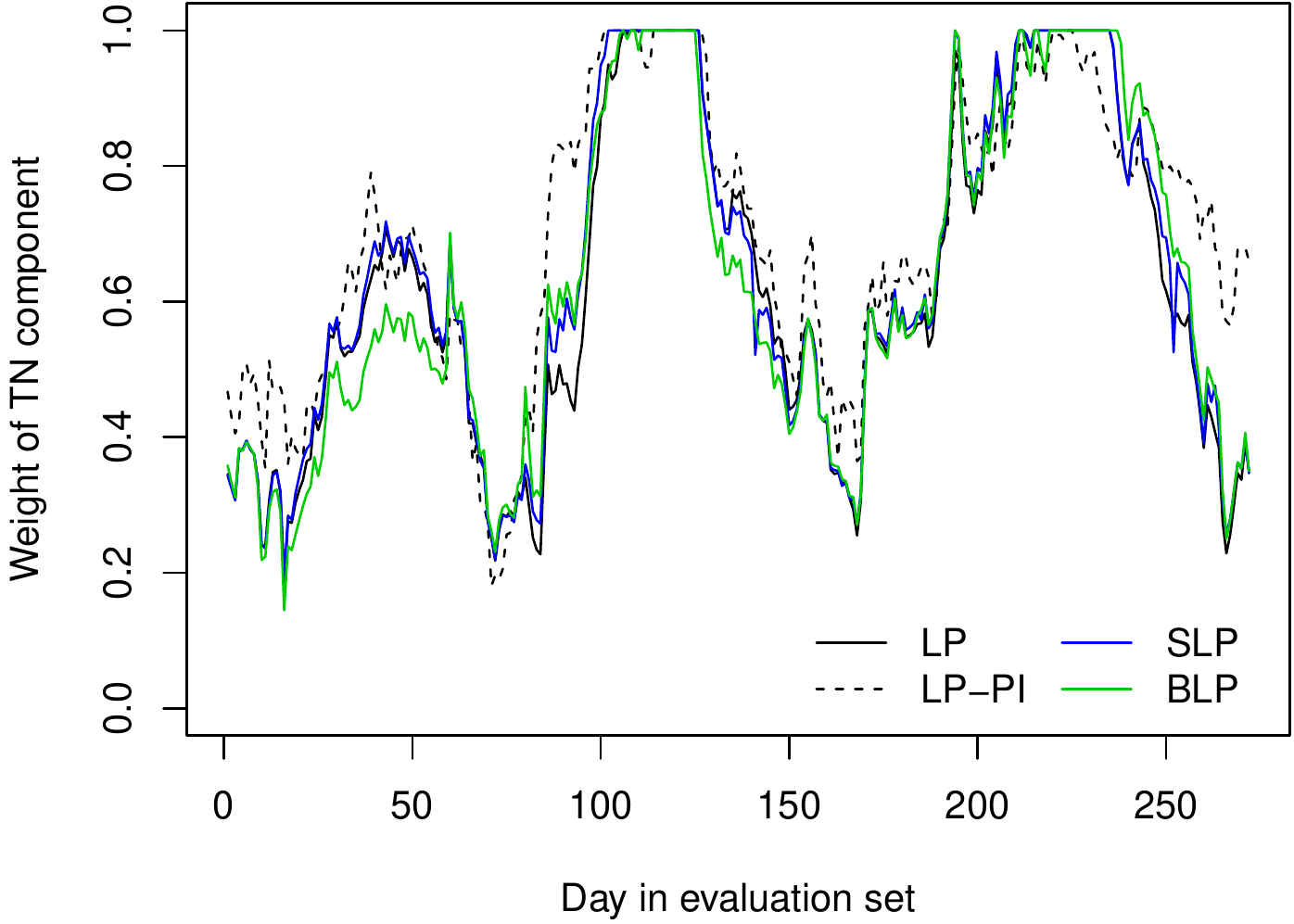} &
\includegraphics[width=0.48\textwidth]{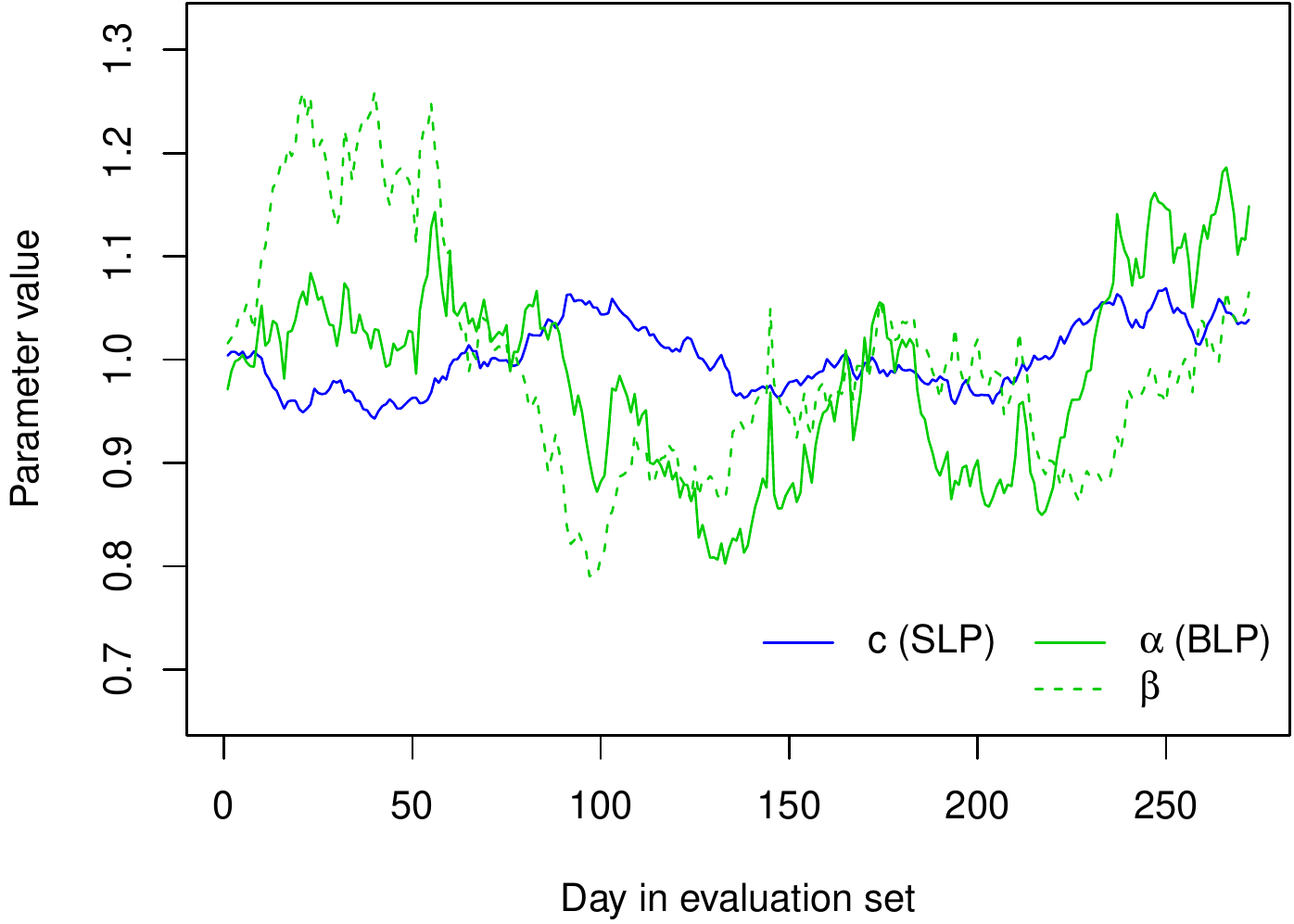} 
\end{tabular}
\caption{Illustration of mixture weights and other combination parameters for the LP, LP-PI, SLP and BLP combination methods over the corresponding verification periods for both wind speed data sets. \label{fig:paramplot-wind}}
\end{figure}

The post-processing models introduced in Section \ref{sec:sec3} are estimated using the optimum score estimation approach described in Section \ref{subs3.3}. The TN and LN component models are estimated by minimizing the mean CRPS over the training sets, whereas ML estimation is employed for the full mixture model \eqref{eq:eq3.3}. Following previous work \citep{bl15, bl16}, we use rolling training periods of length 30 days (UWME data) and 43 days (ALADIN-HUNEPS data), and estimate the parameters regionally by combining forecast cases from all available observation stations to form a single training set for all stations. Note that alternative similarity-based semi-local approaches to selecting the training sets have been investigated in \citet{lb16}.

Given the estimated coefficients of the component models, the combination parameters in the two-step combination approaches are estimated over the corresponding rolling training periods as described in Section \ref{sec:sec-new}. For the UWME data, forecast cases from calendar year 2007 were used to obtain training periods of equal length for all models which are validated on the data of calendar year 2008. For the ALADIN-HUNEPS data, the first 43 days are not included in the evaluation period (27 June 2012 -- 31 March 2013) in order to compare all models over equal training periods. 

Figure \ref{fig:paramplot-wind} graphically illustrates the resulting combination parameters for the LP, LP-PI, SLP and BLP methods. Here, the BMC method is excluded as the parameters vary over the random draws of the algorithm and do not allow for a straightforward summary. For both data sets, the estimated weight parameters are generally very similar for all methods, with minor deviations for the LP-PI and BLP approaches. The spread-adjustment parameter $\ c\ $ in the SLP method does not vary much over time, whereas the $\ \alpha,\beta\ $ parameters in the BLP approach fluctuate much more rapidly.

\begin{table}[t]
\begin{center}
\caption{Mean CRPS for probabilistic wind speed forecasts of the raw ensemble, the TN, LN and TN-LN (ML) EMOS models, and the forecast combination approaches.} \label{tab:crps-wind} 
\medskip
\begin{tabular}{lcc}
\hline
Forecast 	& UWME & ALADIN-HUNEPS \\
\hline 
Ensemble	& 1.353 & 0.804 \\
TN 			& 1.114 & 0.735 \\
LN 			& 1.113 & 0.740 \\
TN-LN (ML) 	& 1.100 & 0.731 \\
LP 			& 1.111 & 0.735 \\
LP-PI 		& 1.111 & 0.735 \\
SLP 		& 1.111 & 0.737 \\
BLP 		& 1.110 & 0.738 \\
BMC 		& 1.106 & 0.738 \\
\hline 
\end{tabular} 
\end{center}
\end{table}

Table \ref{tab:crps-wind} shows the mean CRPS values for all post-processing models and combination approaches for both data sets. All post-processing and combination methods substantially improve the raw ensemble forecasts. Among the post-processing models, the full TN-LN mixture model performs best, and the ranking of the TN and LN model depends on the data set at hand. For the UWME data, all forecast combination methods outperform the individual TN and LN component models, but are unable to compete with the TN-LN mixture model. The relative differences between the combination approaches are small, with the BLP and BMC approaches showing slightly better results. By contrast, none of the combination methods is able to perform better than the TN EMOS model for the ALADIN-HUNEPS data, and the SLP, BLP and BMC approaches result in slightly worse forecasts. Note that the BLP and SLP methods result in worse forecasts compared to the LP approach even though the latter arises as a special case for $\ \alpha = \beta = 1\ $ and $\ c = 1.\ $ A potential explanation for these observations is the danger of over-fitting in choosing the optimal combination parameter values in the training sample that might not be optimal for the corresponding out of sample evaluation set. Further, the ALADIN-HUNEPS data set is comprised of only 10 observation stations. The training sets thus contain fewer forecast cases compared to the UWME data which might favor combination methods with a lower number of parameters.

To assess the variability of the observed score differences and the statistical significance of these findings, we utilize moving block bootstrap resampling \citep{bbs} and Diebold-Mariano \citep[DM;][]{dm95} tests described in the following. Both approaches allow to account for the temporal dependencies in the forecast errors. For a pair of forecast methods $\ F_1,F_2,\ $ denote the vector of CRPS differences by $\ (d_1,\dots,d_n),\ $ with
$$
d_i(F_1, F_2) = \crps(F_1^{(i)}, x_i) - \crps(F_2^{(i)}, x_i),
$$
where $\ F_j^{(i)}\ $  denotes the forecast distribution $\ F_j, j = 1,2,\ $ for forecast case $\ i = 1,\dots,n\ $ in the evaluation set, and $\ x_i\ $ denotes the corresponding observation.

The moving block bootstrap resampling proceeds as follows: Randomly draw a starting date $\ t \in \{1,\dots,T-b+1\},\ $ where $\ T\ $ denotes the number of days in the evaluation set, and $\ b\ $ is the block length. Then select all entries of $\ (d_1,\dots,d_n)\ $ that correspond to any of the  $\ b\ $ consecutive days starting at $\ t,\ $ i.e.~select entries from all observation stations made at days $\ t,t+1,\dots,t+b-1\ $ and compute the mean value of this subset of  $\ (d_1,\dots,d_n).\ $ This procedure is repeated $\ M\ $ times, and we subsequently assess the proportion of bootstrap resampling repetitions with negative mean score differences indicating a superior predictive performance of $\ F_1.\ $ 

\begin{figure}[t]
\begin{tabular}{cc}
(a) & (b) \\[0.5em]
\includegraphics[width=0.48\textwidth]{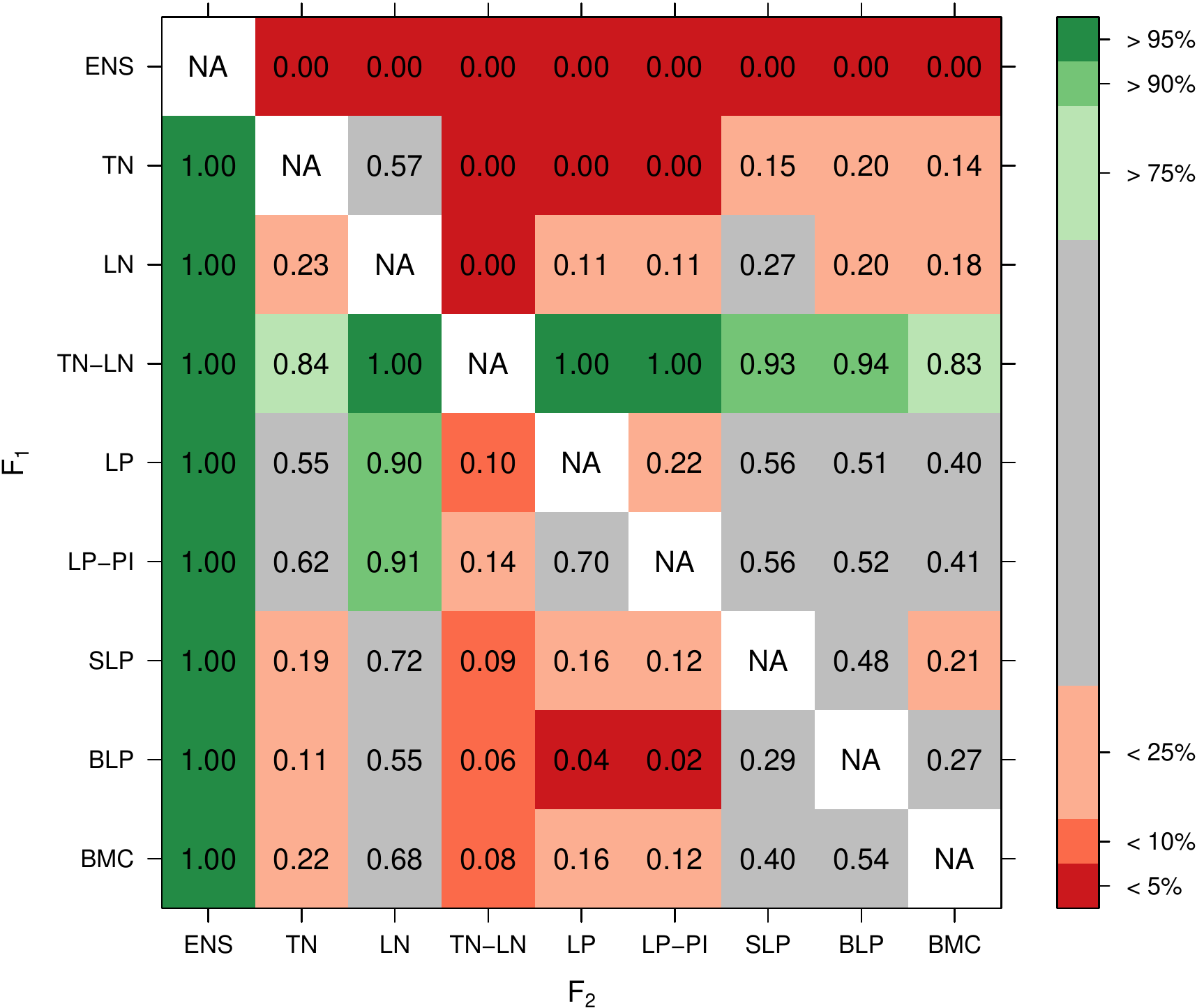} &
\includegraphics[width=0.48\textwidth]{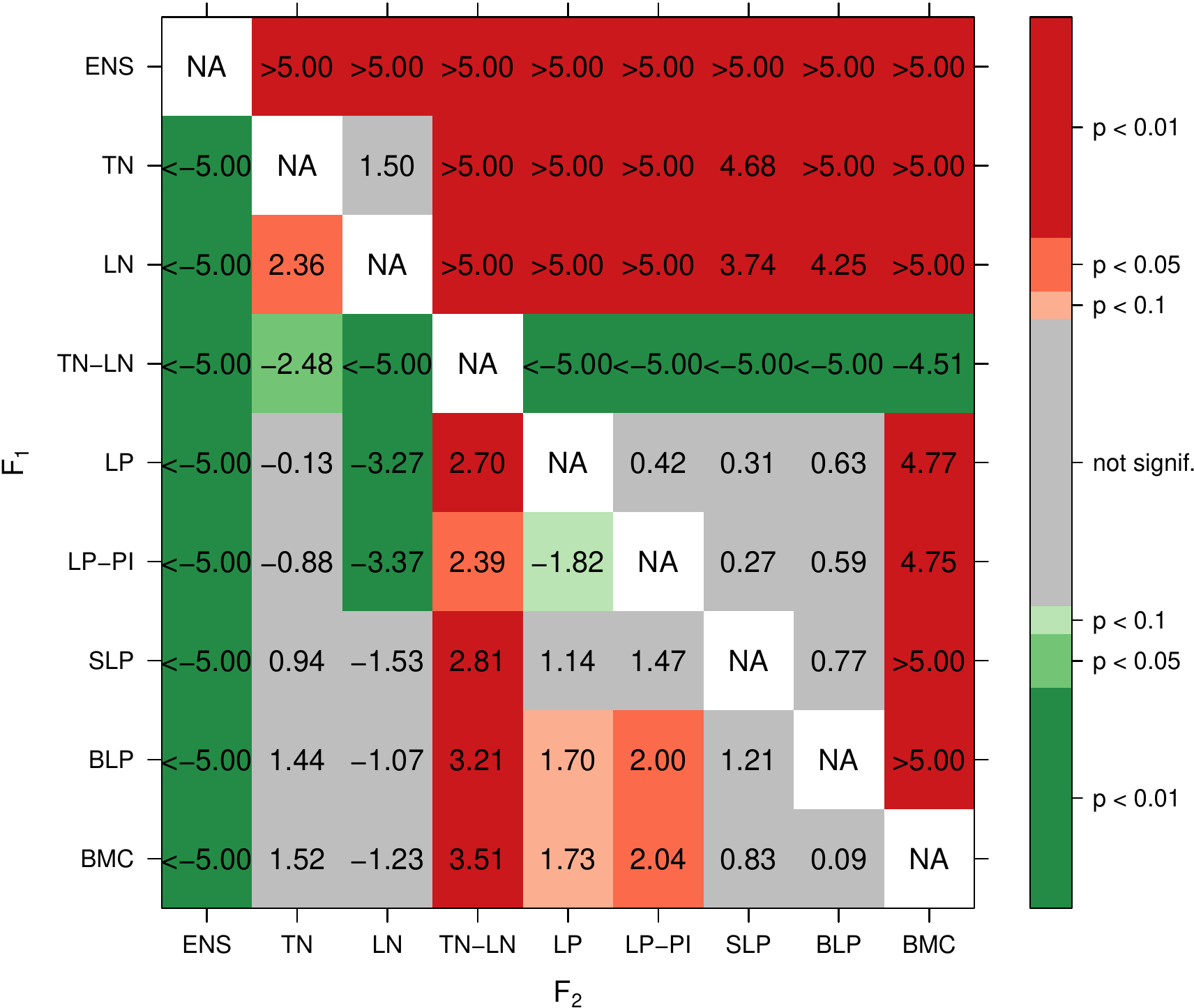}
\end{tabular}
\caption{Summary of (a) block bootstrap resampling and (b) DM test results for both wind speed data sets and all pair-wise comparisons of forecasts. In both plots, the upper triangle contains results for the UWME data, and the lower triangle contains results for the ALADIN-HUNEPS data. In (a), the entry in row $i$ and column $j$ contains the proportion of bootstrap repetitions with negative mean score differences between $F_1$ and $F_2$,  where $F_1$ is the forecast of the model in the $i$-th row, and $F_2$ is the forecast of the model in the $j$-th column, color-coded so that green (red) entries indicate superior performance of the model in the corresponding row (column). Similarly,  values of the DM test statistic $t_n$ are shown for comparisons of $F_1$ and $F_2$ in (b). The values of $t_n$ are color-coded by the corresponding $p$-values of the test statistic under the null hypothesis of equal predictive performance. \label{fig:wind}}
\end{figure}

Figure \ref{fig:wind}(a) graphically summarizes the results of the block bootstrap resampling with a block length of \ $b = 50$ \ days, and \ $M = 10\,000$ \ repetitions. Note that the size of the individual bootstrap samples thus differs for the two data sets due to the different number of observation stations. For the UWME data illustrated in the upper triangle, the differences between the component models and the combination approaches are clearly visible and much more pronounced compared to the differences among the combination approaches where no clear trend can be detected. For the ALADIN-HUNEPS data (lower triangle), the combination methods exhibit some differences. In particular, the BLP approach performs considerably worse compared to the LP and LP-PI methods. For both data sets, the superiority of the full TN-LN mixture model forecasts is clearly visible.  

DM tests are formal statistical tests of equal predictive performance based on the test statistic
\[
t_n = \sqrt{n}\,\frac{\bar d}{\hat{\sigma}_d},
\]
where $\ \bar d = \frac{1}{n}\sum_{i = 1}^{n} d_i\ $ and $\ \hat{\sigma}_d\ $ is an estimator of the asymptotic variance of the score difference. Under standard regularity conditions, $\ t_n\ $ is asymptotically standard normal under the null hypothesis of equal predictive performance of $\ F_1\ $ and $\ F_2.\ $ Negative values of $\ t_n\ $ indicate superior predictive performance of $\ F_1,\ $ and $\ F_2\ $ is preferred if $\ t_n\ $ is positive. As an estimator $\ \hat{\sigma}_d^2\ $, we use the sample autocovariance up to lag
$\ \tau-1\ $ in case of $\ \tau\ $ step ahead forecasts, see \citet{bl16} for details. 

The corresponding results are summarized in Figure \ref{fig:wind}(b). Similar to the results of the block bootstrap resampling, the DM tests reveal a high level of significance of the score differences between the component models and the forecast combination approaches for the UWME data (upper triangle). By contrast, the score differences among the combination methods are not significant. Similar results can be observed for the ALADIN-HUNEPS data (lower triangle), where, however, the differences between the component and combined models are generally smaller and less significant. The only significant score differences among the combination methods are observed between BLP and the approaches based on the linear pool.

\begin{figure}[p]
\includegraphics[width=\textwidth]{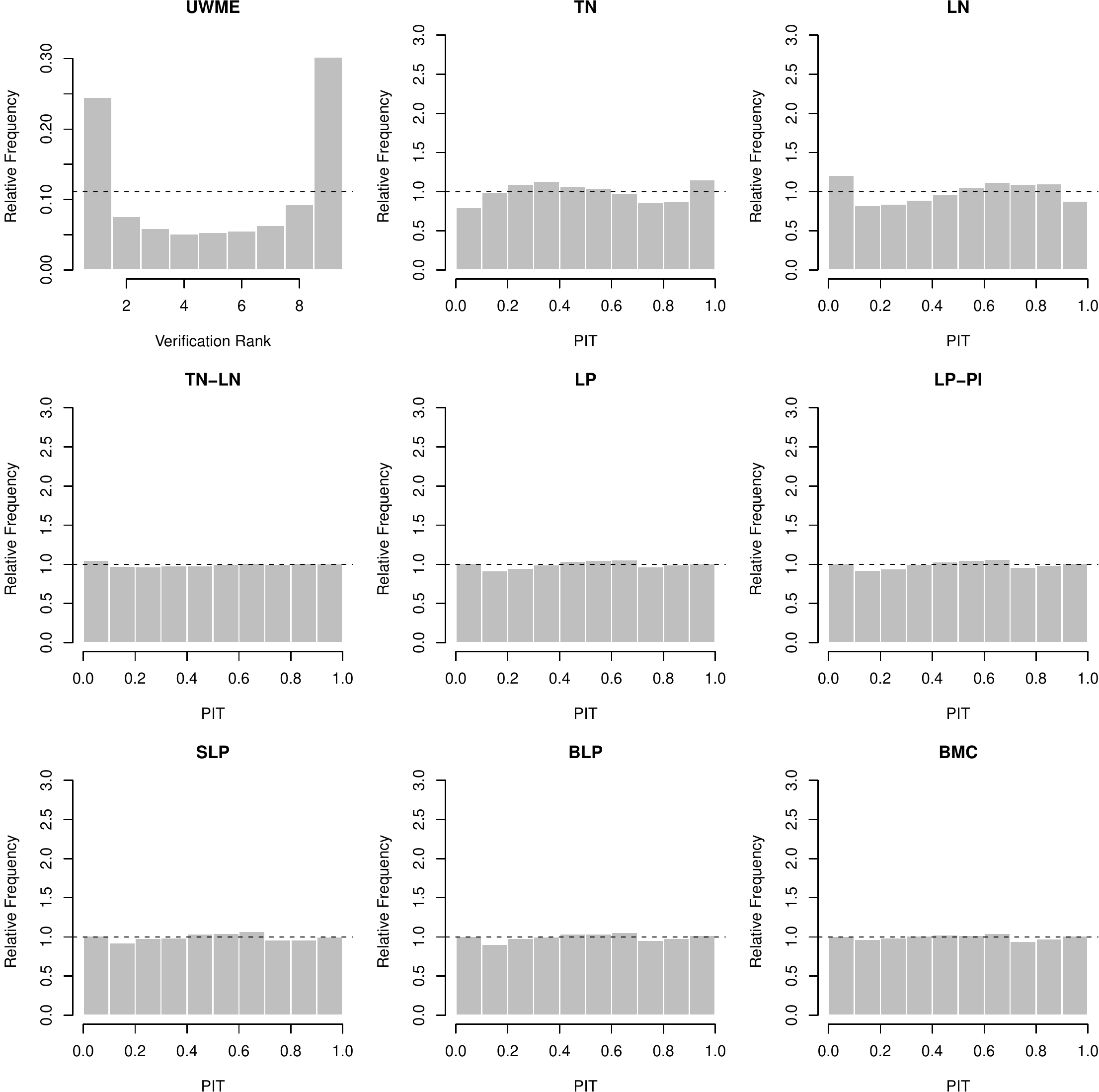}
\caption{Verification rank histogram of raw ensemble forecasts and PIT histograms for post-processed and combined forecast distributions for the UWME data. \label{fig:pits_wind_uwme}}
\end{figure}

The improved predictive performance of the forecast combination approaches compared to the individual EMOS models based on single parametric distributions can be partially explained by the improved calibration of the predictive distributions that will be demonstrated in the following. Calibration of the raw ensemble and the post-processed forecasts can be assessed graphically with the help of verification rank and probability integral transform (PIT) histograms, respectively. The former is the histogram of ranks of the validating observations with respect to the corresponding ensemble predictions computed for all forecast cases \citep[see e.g.][Section 7.7.2]{wilks}. For a calibrated ensemble, the observations and the ensemble forecasts should be exchangeable, resulting in a uniform verification rank histogram. The PIT is  the value of the predictive CDF evaluated at the verifying observation \citep{rgbp}, PIT histograms can therefore be seen as continuous counterparts of verification rank histograms. The visual inspection of deviations from the desired uniform distribution of the verification ranks and PIT values allows to further detect possible reasons of miscalibration \citep{gbr}.

A verification rank histogram of the raw ensemble forecast and PIT histograms of the post-processed and combined forecast distributions for the UWME data are shown in Figure~\ref{fig:pits_wind_uwme}.  Note that for the BMC forecasts PIT values are calculated for all 50 predictive CDFs of a given forecast case.
Compared to the U-shaped verification rank histogram of the underdispersive raw ensemble forecasts where the observation takes too many high and low ranks, all post-processing approaches are better calibrated which is indicated by smaller deviations from the desired uniform distribution of the PIT values. The calibration of the individual TN and LN component models is not perfect, with the TN model showing systematic over-predictions of high wind speeds, and the LN model over-predicting low wind speed values. By contrast, all forecast combination approaches are able to correct for these deficiencies and are well calibrated, similar to the full TN-LN mixture model. The differences in calibration among the combination methods are small, as the PIT histograms are virtually indistinguishable. The results for the ALADIN-HUNEPS data are qualitatively similar, the corresponding Figure \ref{fig:pits_wind_alhu} is shown in Appendix \ref{sec:appendix}.

\subsection{Precipitation}

Similar to wind speed, the post-processing models introduced in Section \ref{sec:sec3} are estimated using optimum score estimation approaches. The coefficients of the EMOS models \eqref{eq:eq3.4} and \eqref{eq:eq3.6} based on single CSG and GEV distributions are obtained using rolling training periods of lengths 70 (UWME) and 55 days (ALADIN-HUNEPS), which ensures comparability with \citet{bn}.

Given the estimated coefficients of the CSG and GEV component models, the parameters of the two-step combination approaches are estimated as described for wind speed. For the UWME data, forecast cases from calendar year 2007 are again used to obtain training periods of equal length for all models, whereas the first 55 days are excluded from the evaluation period of the ALADIN-HUNEPS data. In this way UWME forecasts are again validated on data from calendar year 2008, whereas the verification period for ALADIN-HUNEPS precipitation forecasts is 21 January -- 25 March 2011.

\begin{figure}[p]
\begin{tabular}{cc}
\multicolumn{2}{c}{UWME} \\[0.5em]
(a) combination weight & (b) further parameters \\[0.5em]
\includegraphics[width=0.48\textwidth]{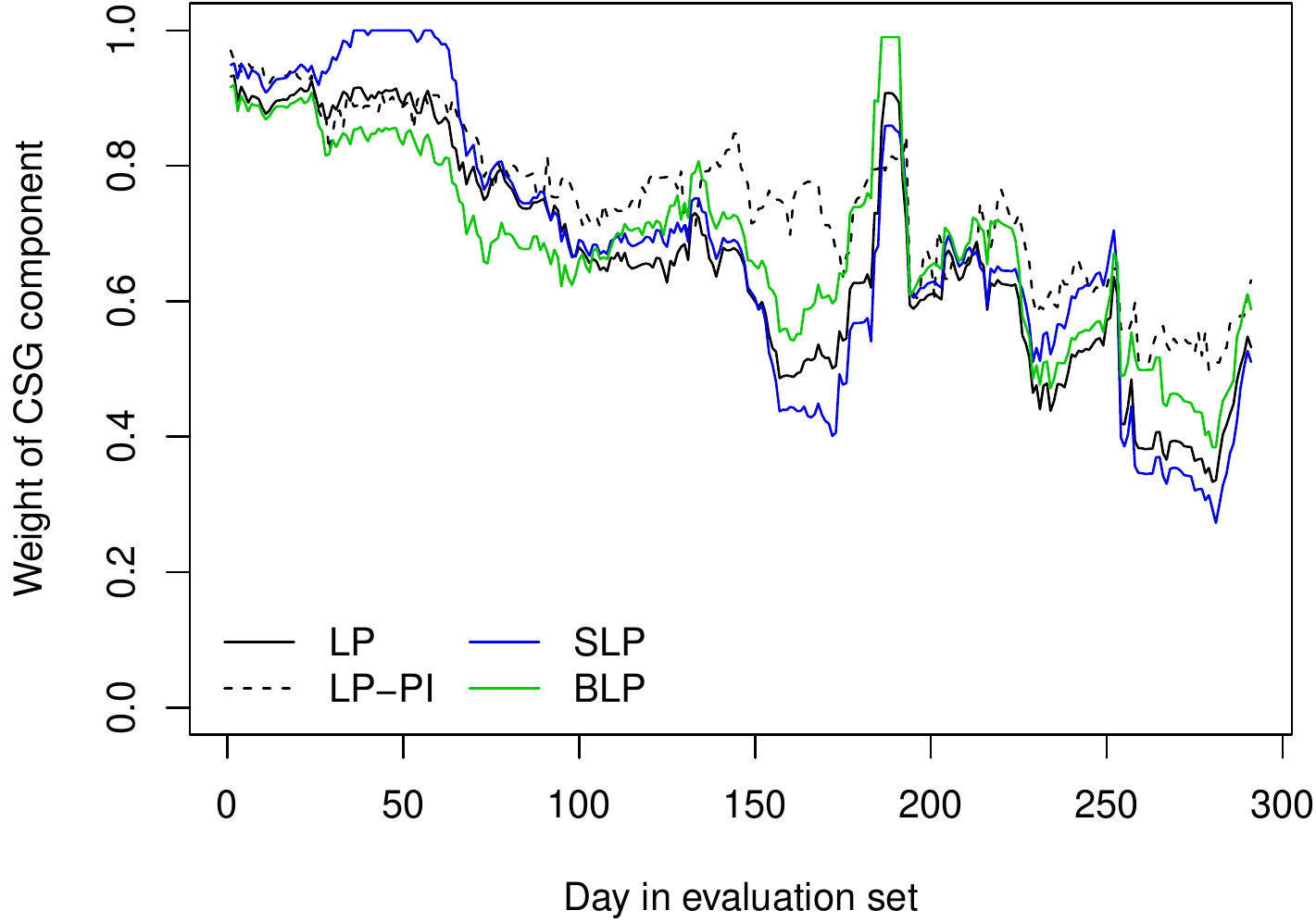} &
\includegraphics[width=0.48\textwidth]{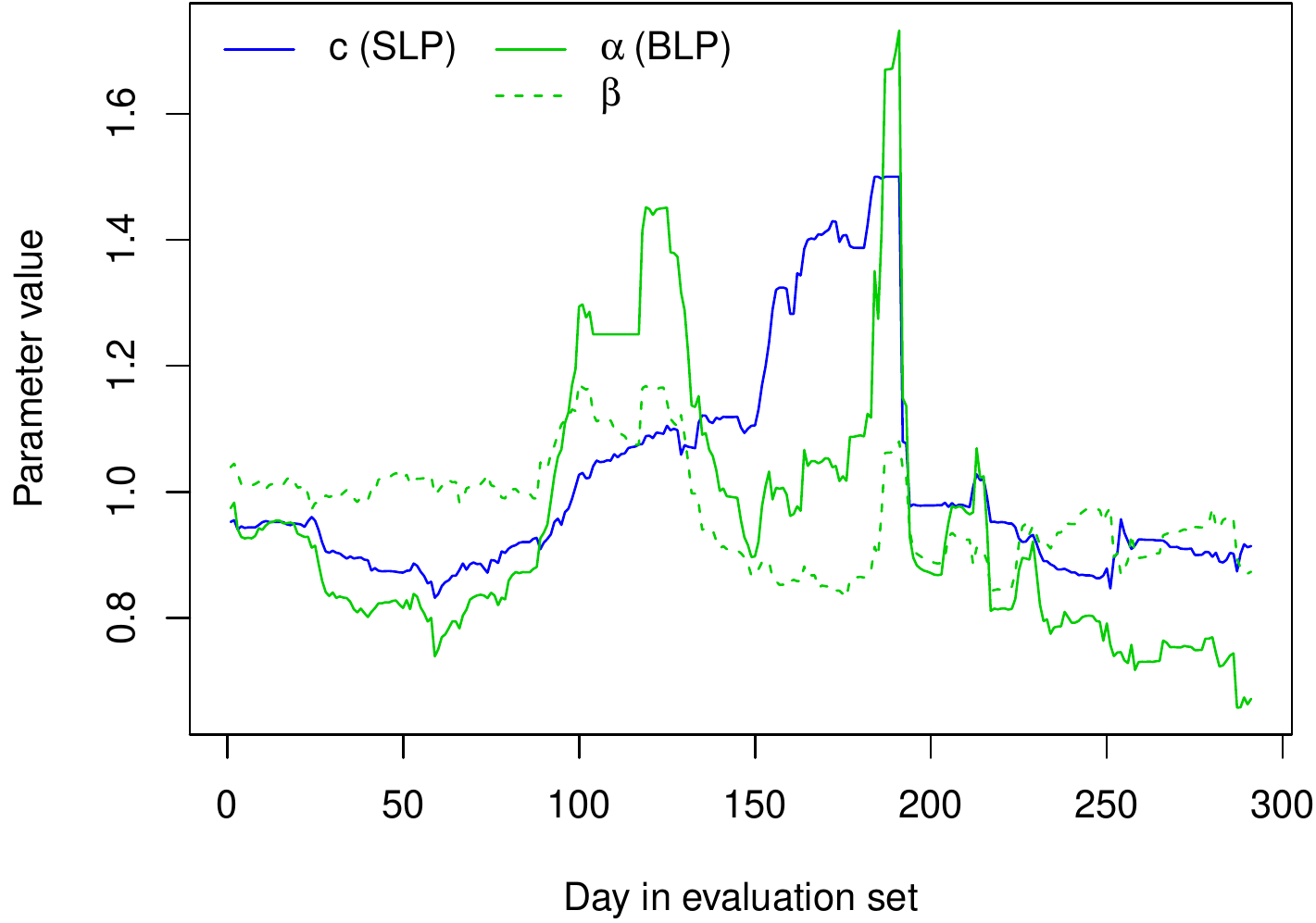} \\[1em]
\multicolumn{2}{c}{ALADIN-HUNEPS} \\[0.5em]
(c) combination weight & (d) further parameters \\[0.5em]
\includegraphics[width=0.48\textwidth]{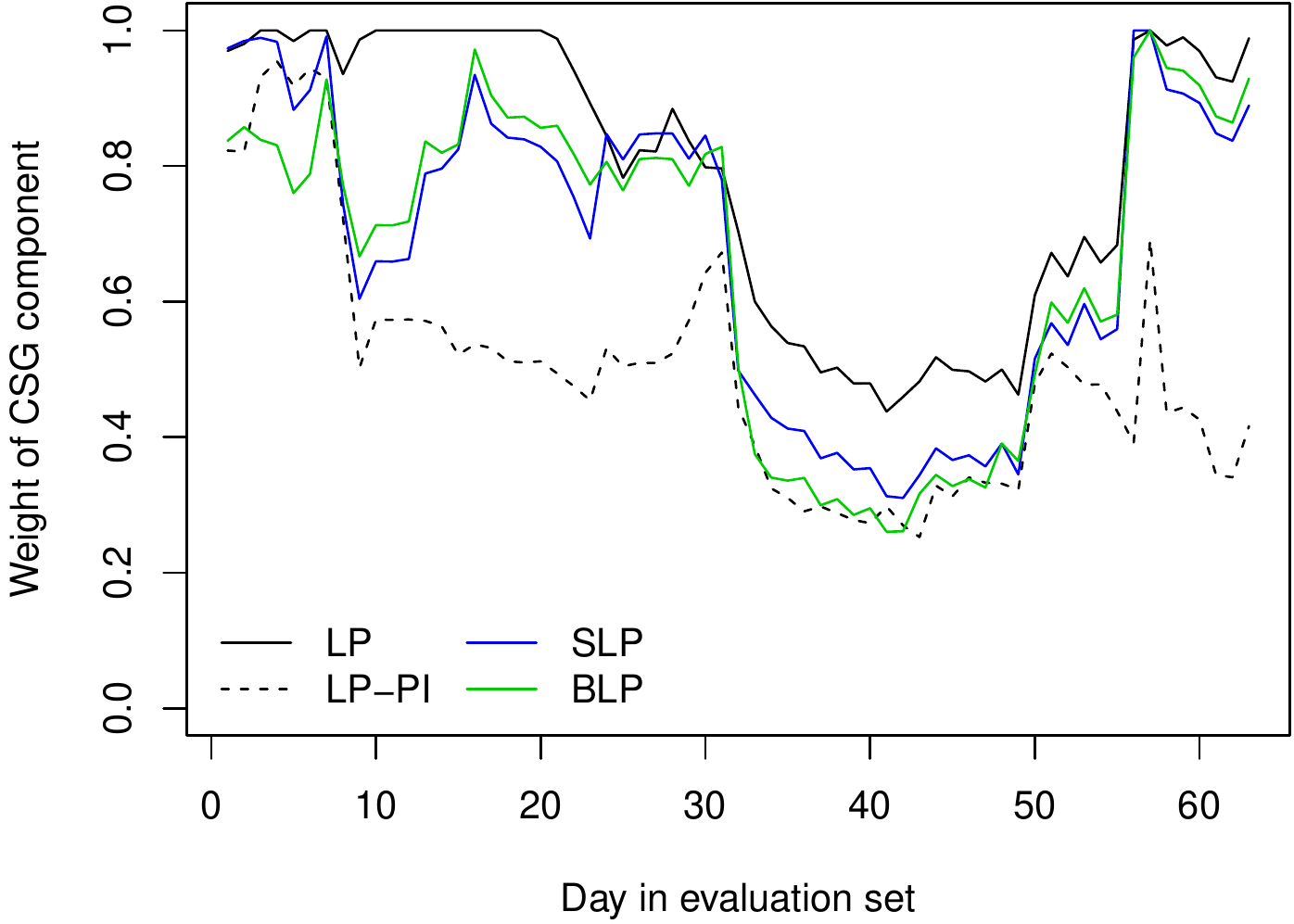} &
\includegraphics[width=0.48\textwidth]{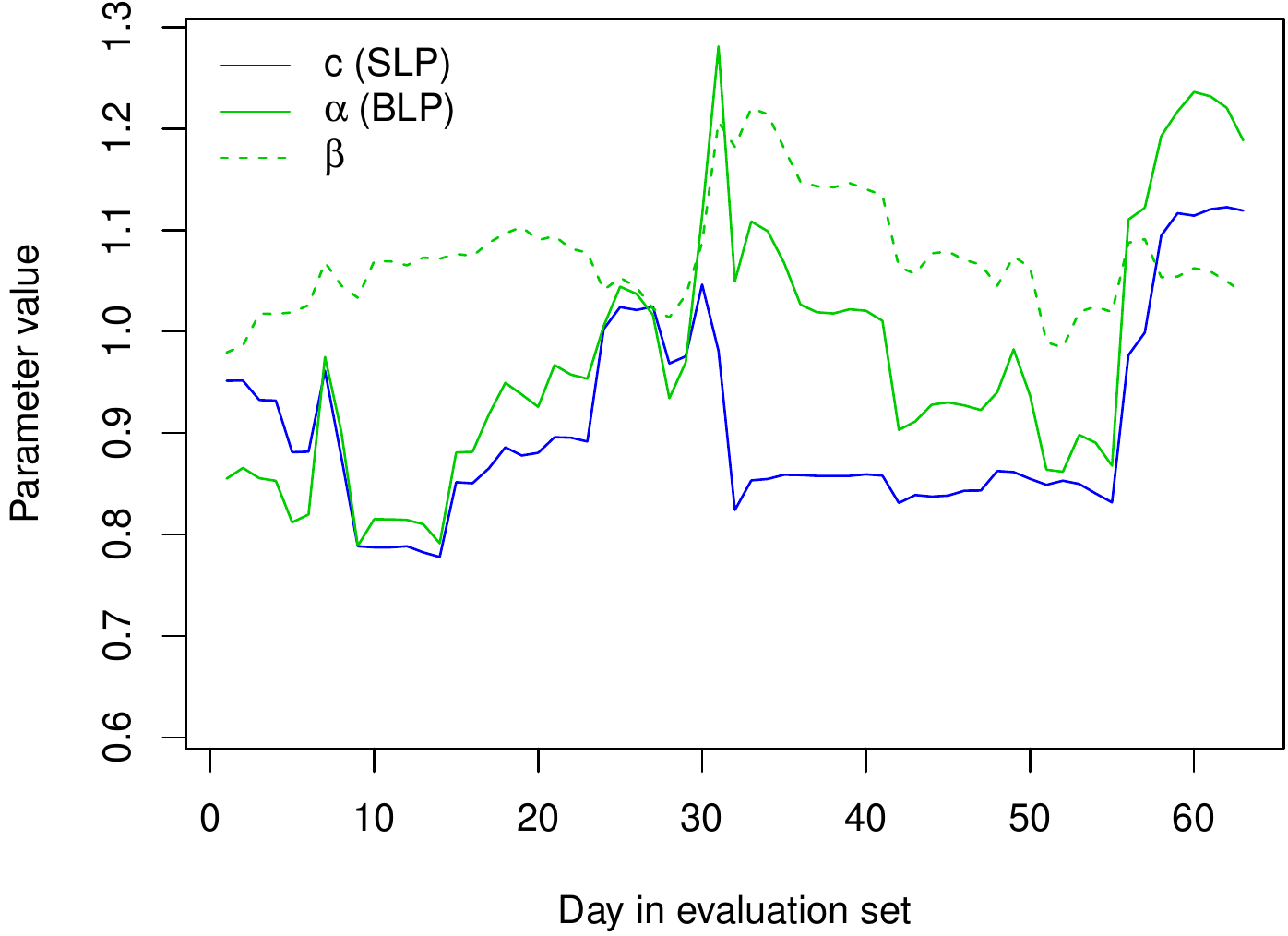}
\end{tabular}
\caption{Illustration of mixture weights and other combination parameters for the LP, LP-PI, SLP and BLP combination methods over the corresponding verification periods for both precipitation data sets. \label{fig:paramplot-prec}}
\end{figure}

The estimates of the combination parameters over the evaluation period are shown in Figure \ref{fig:paramplot-prec}. The mixture weights of the LP, LP-PI, SLP  and BLP approaches exhibit relatively similar developments over time, and the spread-adjustment parameter $\ c\ $ shows slightly higher variability compared to the wind speed forecasts.

\begin{table}
\begin{center}
\caption{Mean CRPS for probabilistic precipitation accumulation forecasts of the raw ensemble, the CSG and GEV EMOS models, and the forecast combination approaches.} \label{tab:crps-prec} 
\medskip
\begin{tabular}{lcc}
\hline
Forecast & UWME & ALADIN-HUNEPS \\
\hline 
Ensemble & 2.884  & 0.269  \\
CSG &  2.198 & 0.258 \\
GEV & 2.227 & 0.264 \\
LP & 2.189 & 0.259 \\
LP-PI & 2.189  & 0.262 \\
SLP & 2.190 & 0.263  \\
BLP & 2.184 & 0.261 \\
BMC & 2.288 & 0.267 \\
\hline 
\end{tabular} 
\end{center}
\end{table}

Mean CRPS values for all post-processing models and forecast combination methods for both data sets are shown in Table \ref{tab:crps-prec}. Compared to wind speed, the relative improvements of both EMOS models over the raw ensemble forecasts are smaller, particularly for the ALADIN-HUNEPS data. This observation is in line with various comparative studies of post-processing models for different variables \citep[see for example][]{HemriEtAl2014}. The CSG model outperforms the GEV model, and in case of the UWME data, the predictive performance is further improved by combining the forecasts via the LP, LP-PI, SLP and BLP approaches which show small relative differences. In light of the larger relative score differences in favor of the CSG method it is worth noting that the estimated mixture weights of the CSG component in these combination approaches are between 0.3 and 0.7 for a large number of forecast cases, see Figure \ref{fig:paramplot-prec}. As observed for wind speed, none of the combination methods is able to outperform the best component model for the ALADIN-HUNEPS data. Forecasts produced by the BMC method are worse than both component models for both data sets and only marginally better than the raw ensemble forecasts in case of the ALADIN-HUNEPS data.

\begin{figure}
\begin{tabular}{cc}
(a) & (b) \\[0.5em]
\includegraphics[width=0.49\textwidth]{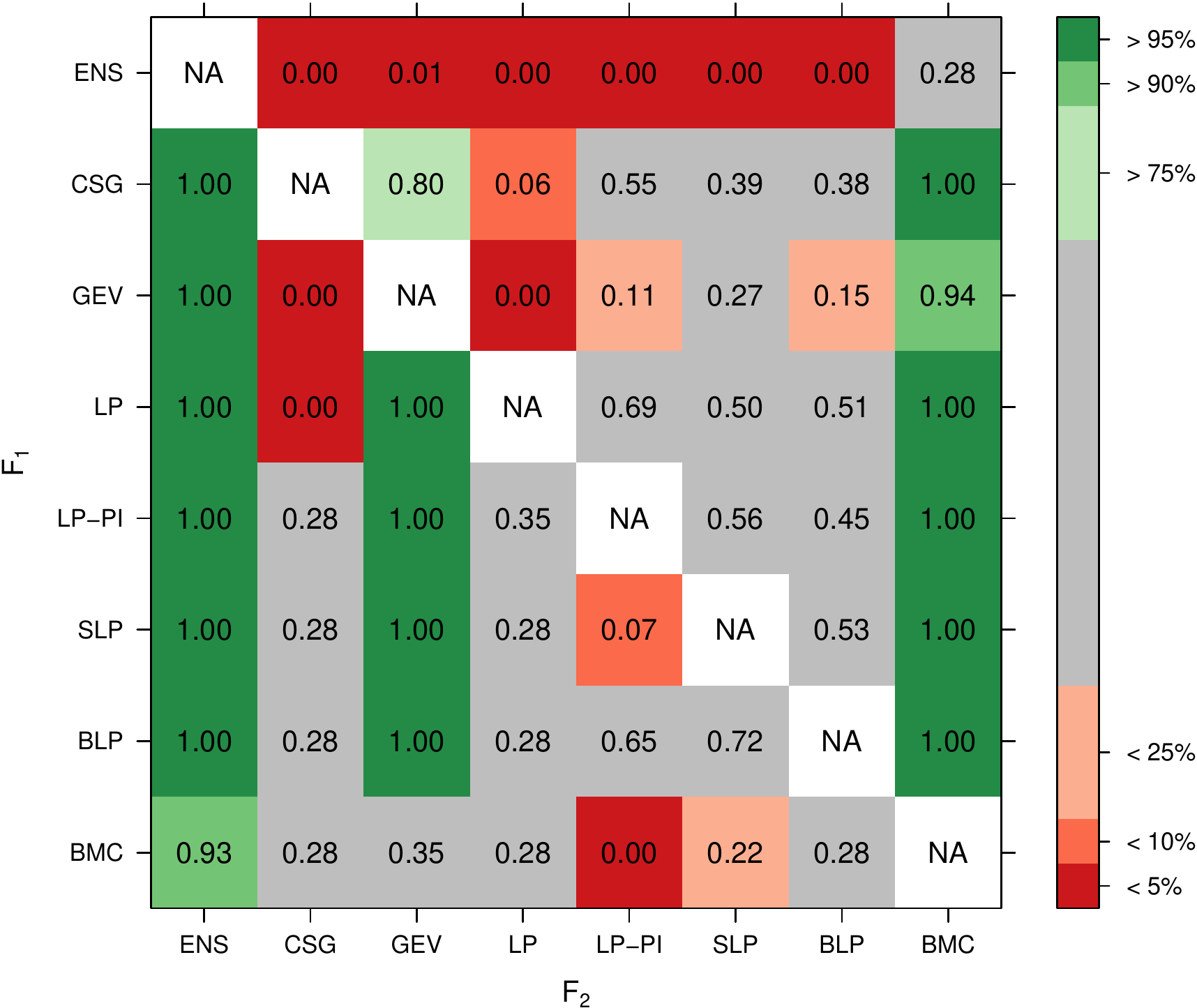} &
\includegraphics[width=0.49\textwidth]{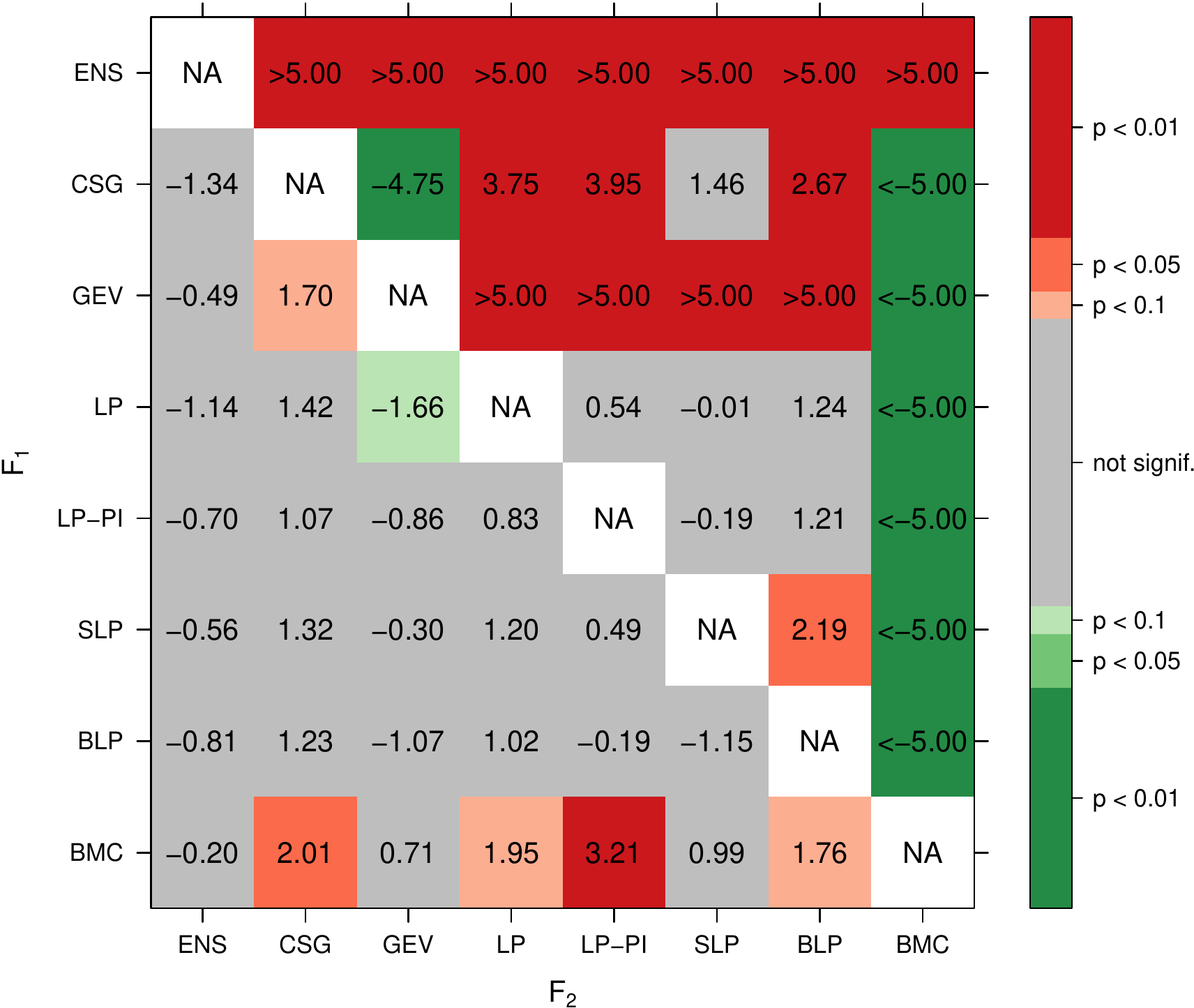}
\end{tabular}
\caption{Summary of (a) block bootstrap resampling and (b) DM test results for both data sets and all pair-wise comparisons of forecasts similar to Figure \ref{fig:wind}, but for precipitation accumulation. In both plots, the upper triangle contains results for the UWME data, and the lower triangle contains results for the ALADIN-HUNEPS data. \label{fig:prec}}
\end{figure}

The variability and statistical significance of the observed score differences is again assessed using moving block bootstrap resampling and DM tests with the setup described above. The results are summarized in Figure \ref{fig:prec}. For the ALADIN-HUNEPS data, the block bootstrap resampling indicates clear differences between the raw ensemble forecasts and all post-processing and forecast combination approaches. Further, the GEV model forecasts are substantially improved by the LP, LP-PI, SLP and BLP combination methods. These differences are less pronounced in the DM tests where none of the score differences show a high level of significance. For the UWME data, the differences between the individual post-processing models and the forecast combination approaches are more pronounced, specifically in terms of the DM tests. The score differences between the LP, LP-PI, SLP and BLP approaches are not significant, however, as expected the BMC forecasts perform substantially worse compared to all alternatives for both data sets.

In contrast to the wind speed forecast discussed above, the EMOS models for precipitation accumulation provide relatively well calibrated forecasts for both data sets, and the forecast combination methods only result in slightly improved calibration. The corresponding verification rank and PIT histograms are provided in Figures \ref{fig:pits-prec_uwme}  and \ref{fig:pits-prec_alhu} in Appendix \ref{sec:appendix}.

\section{Conclusions}
  \label{sec:sec5}

We have investigated the feasibility of using forecast combination approaches to improve the predictive performance of statistical post-processing models based on single parametric families. In general, the results strongly depend on the data set at hand, and forecast combination may either provide slight improvements, or even result in worse forecasts compared to the best mixture component. However, none of the forecast combination methods is able to outperform a jointly estimated full TN-LN mixture EMOS models for wind speed. 

The relative improvements obtained through forecast combination are larger in the case of wind speed where the PIT histograms reveal complementary systematic errors in calibration. All combination approaches result in calibrated forecast distributions and are thus able to correct the over-prediction of high (LN) or low (TN) wind speed values. The differences among the combination approaches are generally small, except for precipitation accumulation where the BMC method leads to considerably worse forecasts. The provisional adaptation described in Section \ref{sec:bmc} fixes problems that occurred in a naive application of the BMC sampling algorithm. However, the worse results compared to the competing combination methods call for an extension of the methodology towards forecast distributions with point masses that is beyond the scope of the present paper.

Compared to previous work of \citet{MoellerGross2016} and \citet{BassettiEtAl2015}, we generally do not find as substantial differences in predictive performance between the individual EMOS models and the combination approaches. Further, compared to the case study of wind speed data in \citet{BassettiEtAl2015} the LP and BMC method show much less significant differences. It might thus be interesting to investigate which features of the data and mixture component models are beneficial for which of the combination approaches.

The larger relative improvements of forecast combination for the UWME data sets may indicate that longer training periods are generally better suited to determine the combination parameters. Therefore, training sets expanding with time might improve the predictive performance. Further, local estimation of the combination parameters using only data from the single observation station of interest, or alternative similarity-based semilocal approaches \citep{lb16} might be better able to account for locally varying station-dependent features of the forecast errors of the individual EMOS models. Here we used equal regional training periods for all models to ensure direct comparability, and leave such extensions for future work.

The BLP and SLP methods result in worse forecasts compared to the LP approach for both ALADIN-HUNEPS data sets even though the latter arises as a special case. As discussed above, this is likely due to  over-fitting in choosing the optimal combination parameter values in the training sample that might not be optimal for the out of sample evaluation period. The LP method may be more robust in this sense due to the smaller number of estimated parameters.

We have proposed a plug-in variant of the linear pool that utilizes the most recent EMOS coefficient estimates to replace those in the training period used to determine the mixture weight, and thereby reduces the number of required numerical integrations to a single one. Despite discarding potentially useful information for forecast combination, the LP-PI approach results in very similar mixture weights and predictive performance compared to the traditional linear pool, and may thus offer an alternative option if the computational costs of estimating the mixture weight are high.

Forecast combination methods may also offer a new approach to post-processing multi-model ensemble predictions such as the TIGGE forecasts \citep{tigge16}. Instead of utilizing forecasts from all models as input of an EMOS model based on a single parametric distribution it might be helpful to post-process the ensemble predictions of the different models independently, and then subsequently combine the forecast distributions with the approaches discussed above. A further interesting starting point for future research might be the application of the combination methods to other weather variables such as total cloud cover \citep{tcc}.

An entirely different approach to post-processing that completely circumvents the problem of choosing suitable parametric forecast distributions is the use of non-parametric methods, see for example \citet{hw06,flowerdew}, and \citet{taillardat}. However, these approaches suffer from the limitation that the support of the forecast distribution is restricted to the range of observed values in the training sets. Further, these methods require sufficiently long training periods, and generally lead to high computational costs.

\bigskip
\noindent
{\bf Acknowledgments.} \ 
The work leading to this paper was done during visits of S\'andor Baran at the Heidelberg Institute for Theoretical Studies in the framework of the visiting scientist program and DAAD program ``Research Stays for University Academics and Scientists, 2017''.
S\'andor Baran was also supported by the J\'anos Bolyai Research Scholarship of the Hungarian Academy of Sciences.
Sebastian Lerch gratefully acknowledges support by Deutsche Forschungsgemeinschaft (DFG) through the project ``C7 -- Statistical post-processing and stochastic physics for ensemble predictions'' within Transregional Collaborative Research Center 165 ``Waves to Weather'', and thanks the Klaus Tschira Foundation for infrastructural support at the Heidelberg Institute for Theoretical Studies. Helpful comments by Tilmann Gneiting are gratefully acknowledged. The authors thank Francesco Ravazzolo for kindly providing a Matlab implementation of the BMC method, and further thank the University of Washington MURI group for providing the UWME data and Mih\'aly Sz\H ucs from the HMS for providing the ALADIN-HUNEPS data.

\appendix

\newpage 
\section{Additional figures}\label{sec:appendix}

\bigbreak\bigbreak\bigbreak

\begin{figure}[h!]
\includegraphics[width=\textwidth]{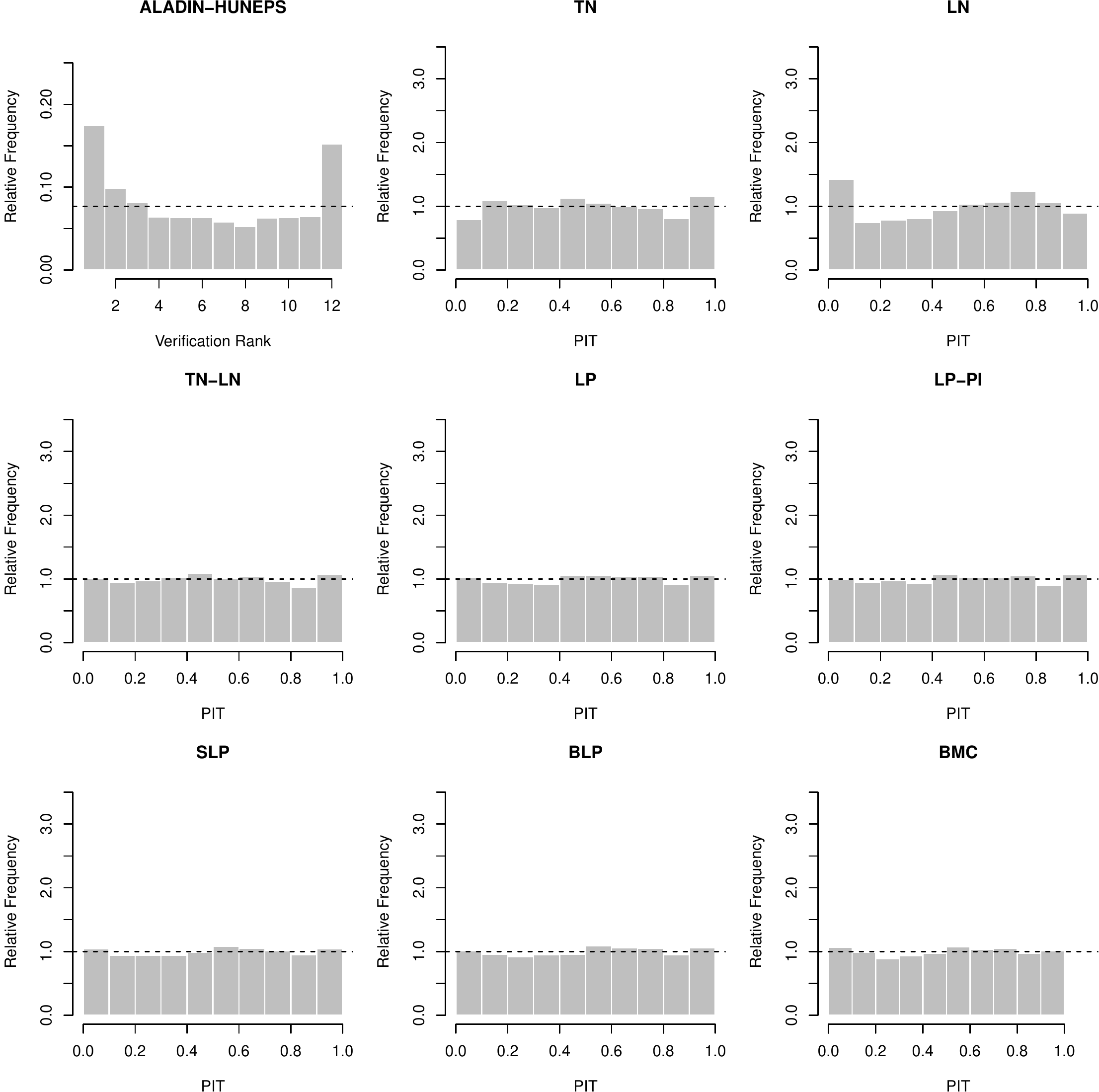}
\caption{Verification rank histogram of raw ensemble forecasts and PIT histograms for post-processed and combined forecast distributions for the ALADIN-HUNEPS wind speed data. \label{fig:pits_wind_alhu}}
\end{figure}

\begin{figure}[p]
\centering
\includegraphics[width=\textwidth]{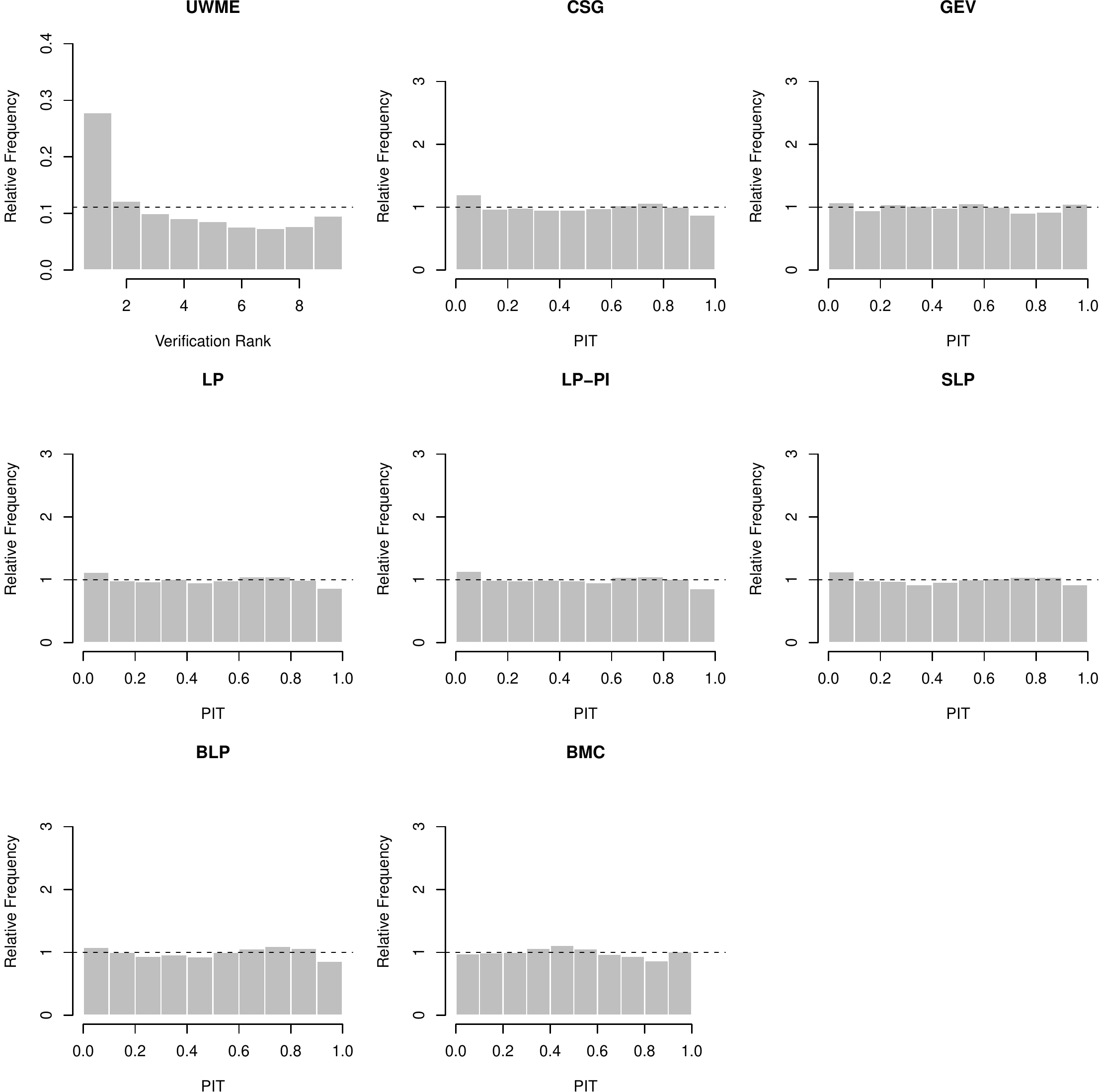}
\caption{Verification rank histogram of raw ensemble forecasts and randomized PIT histograms for post-processed and combined forecast distributions for the UWME precipitation data. Note that to account for the discrete-continuous nature of the models for precipitation accumulation, in case of zero  observed precipitation the PITs are randomized in that a random value is chosen uniformly from the interval between zero and the probability of no precipitation \citep{srgf}. Similarly, in case of precipitation accumulation zero observations are randomized among all zero forecasts to compute the verification rank histograms for the raw ensemble forecasts. \label{fig:pits-prec_uwme}}
\end{figure}

\begin{figure}[p]
\centering
\includegraphics[width=\textwidth]{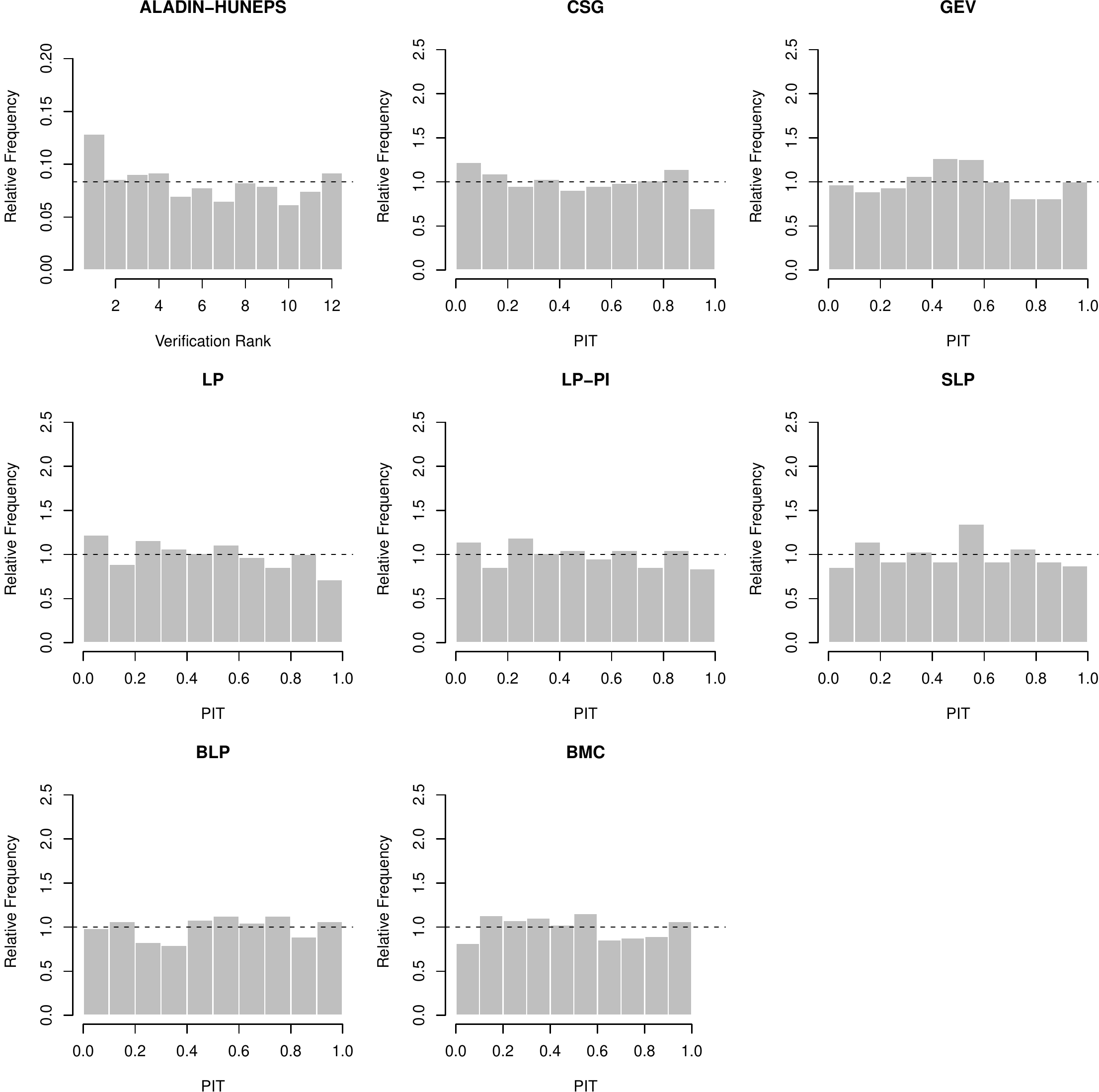}
\caption{Verification rank histogram of raw ensemble forecasts and randomized PIT histograms for post-processed and combined forecast distributions for the ALADIN-HUNEPS precipitation data.  \label{fig:pits-prec_alhu}}
\end{figure}

\end{document}